\documentclass[a4paper,11pt]{article}
\pdfoutput=1 

\usepackage{jcappub} 

\usepackage[T1]{fontenc} 
\usepackage{graphicx}
\usepackage{epsfig}
\usepackage{rotate}
\usepackage{amsmath}
\usepackage{amssymb}
\usepackage{amsfonts}
\usepackage{multirow}
\usepackage{tablefootnote}
\usepackage{diagbox}
\usepackage{slashbox}
\usepackage{bm}
\usepackage{enumerate}
\usepackage{afterpage}
\usepackage{xcolor}
\usepackage{natbib, hyperref}
\usepackage{cancel}

\UseRawInputEncoding

\def\be{\begin{equation}}
\def\ee{\end{equation}}
\def\bea{\begin{eqnarray}}
\def\eea{\end{eqnarray}}
\def\n{\bm{n}}
\def\k{\bm{k}}
\def\v{\bm{v}}

\def\H{{\cal H}}

\def\tw{(2)}

\newcommand{\hi}{{\rm HI}}
\newcommand{\ud}{\mathrm{d}}
\newcommand{\p}{\partial}
\newcommand{\cH}{\mathcal{H}}

\newcommand{\red}[1]{{\color{black}{#1}}}

\title{{Detecting the relativistic bispectrum in 21cm intensity maps}}

\author{Sheean Jolicoeur$^1$, Roy Maartens$^{1,2}$,  Eline M. De Weerd$^3$, \\ Obinna Umeh$^2$, Chris Clarkson$^{3,1}$, Stefano Camera$^{4,5,1}$}
\affiliation{$^1$Department of Physics \& Astronomy, University of the Western Cape, Cape Town 7535, South Africa\\
$^{2}$Institute of Cosmology \& Gravitation, University of Portsmouth, Portsmouth PO1 3FX, UK\\
$^{3}$School of Physics \& Astronomy, Queen Mary University of London, London E1 4NS, UK \\
$^{4}$Dipartimento di Fisica, Universit\`a degli Studi di Torino, 10125 Torino, Italy\\
$^5$Istituto Nazionale di Fisica Nucleare, Sezione di Torino, 10125 Torino, Italy}

\emailAdd{jolicoeursheean@gmail.com, roy.maartens@gmail.com, e.deweerd@qmul.ac.uk, obinna.umeh@port.ac.uk, chris.clarkson@qmul.ac.uk, stefano.camera@unito.it}

\abstract{We investigate the detectability of leading-order relativistic effects in the bispectrum of future 21cm intensity  mapping surveys. The relativistic signal arises from Doppler and other line-of-sight effects in redshift space. In the power spectrum of a single tracer, these effects are suppressed by a factor $\cH^2/k^2$. By contrast, in the bispectrum the relativistic signal couples to short-scale modes, leading to
an imaginary contribution that scales as $\cH/k$, thus increasing the possibility of detection.
Previous work has shown that this relativistic signal is detectable in a Stage IV H$\alpha$ galaxy survey. 
{We show that the signal is also detectable by next-generation  21cm intensity maps, but typically with a lower signal-to-noise, due to foreground and telescope beam effects.}}

\begin{document}
\maketitle
\date{\today}
\flushbottom

\section{Introduction}
{Next-generation galaxy surveys will allow for precision measurements of the  bispectrum, bringing new information to improve constraints on cosmological parameters and to break some of their degeneracies (see e.g. \cite{Scoccimarro:2015bla,Tellarini:2016sgp, Gil-Marin:2016wya,  Slepian:2016kfz,Gagrani:2016rfy, Sugiyama:2018yzo,Desjacques:2018pfv,Child:2018klv, Yankelevich:2018uaz, Schmit:2018rtf,DiDio:2018unb,Gualdi:2019ybt,Sarkar:2019ojl,Chudaykin:2019ock,Oddo:2019run, Sugiyama:2019ike,Philcox:2019hdi,Durrer:2020orn,Montanari:2020uez}). 

These surveys will typically probe very large volumes in the Universe, including ultra-large scales ($k\lesssim k_{\rm eq} \sim 0.01\;\mathrm{Mpc}^{-1}$), which facilitates the detection of, or precision constraints on, primordial non-Gaussianity \cite{Tellarini:2015faa,Watkinson:2017zbs,Majumdar:2017tdm,Karagiannis:2018jdt,Karagiannis:2019jjx,Bharadwaj:2020wkc}
and relativistic effects
\cite{Kehagias:2015tda,Umeh:2016nuh, DiDio:2016gpd, Jolicoeur:2017nyt,Bertacca:2017dzm,Jolicoeur:2017eyi,Koyama:2018ttg,Clarkson:2018dwn,Maartens:2019yhx,Jeong:2019igb}.
 
Redshift-space distortions (RSD) are the dominant observational effect on the number counts or brightness temperature at first and second order in perturbations. There are local relativistic corrections to standard RSD, arising from Doppler and other line-of-sight gradient terms, together with their couplings.  In Fourier space, the leading-order corrections  scale as i\,$(\cH/k)$. In the tree-level power spectrum of a single tracer, the only nonzero contribution from these Doppler-type effects is from their square, since the power spectrum is necessarily real. As a result, the Doppler-type effects are further suppressed in the auto-power spectrum, scaling as  $(\cH/k)^2$. 

The tree-level bispectrum of a single tracer is not forced to be real and it couples Doppler-type effects with the standard (`Newtonian') density + RSD term. Consequently, the leading-order relativistic contribution to the bispectrum scales as i\,$(\cH/k)$. 
This means that the leading-order relativistic signal is more detectable in the bispectrum than the power spectrum, for a single tracer.
In Fourier space, the dominant relativistic effect is a purely imaginary part of the bispectrum, as shown by \cite{Clarkson:2018dwn,Maartens:2019yhx}. Our previous work \cite{Maartens:2019yhx} showed that it is detectable by a Stage IV spectroscopic galaxy survey.

 In this work, we investigate the detectability of the relativistic signal in the bispectrum of various planned 21cm intensity mapping surveys at post-reionisation redshifts. 
The 21cm emission line of neutral hydrogen (HI) is measured without detecting the individual galaxies that contain HI. This results in brightness temperature maps that trace the large-scale structure with exquisite redshift precision. In Section 2 we discuss the leading order relativistic form of the temperature contrast up to second order, and its contribution to the bispectrum. Section 3 describes the signal, modelled using the tree-level bispectrum with the addition of a phenomenological model to account for RSD `fingers-of-god' nonlinearity.
Foreground contamination overwhelms the signal, and cleaning techniques must be applied which lead to a loss of signal in regions of Fourier space, which we take into account. We also discuss the effects of telescope beams and the instrumental noise. Our forecast signal-to-noise for future surveys is presented in Section 4, and we conclude in Section 5.}

\section{{Relativistic effects in the 21cm intensity bispectrum}}

{The HI brightness temperature  measured at redshift $z$ in direction $\n$  is related to the observed number of 21cm emitters per redshift per solid angle, $N_{\hi}$, as follows (see \cite{Hall:2012wd,Alonso:2015uua} for details):
\be
T_{\hi}(z,\n)= {\rm const.}\, { N_{\hi}(z,\n)  \over d_{\rm A}(z,\n)^2}\,,
\label{tobs}
\ee
where $d_{\rm A}$ is the angular diameter distance.  

{The background HI brightness temperature  follows from \eqref{tobs} as \cite{Villaescusa-Navarro:2018vsg}
\be \label{bart}
\bar{T}_{\mathrm{HI}}(z)= 189h\,{(1+z)H_0 \over \H(z)}\Omega_{\hi}(z)~~{\rm mK}.
\ee
Here $h=H_0/(100\,$km/s), $\H=(\ln a)'$ is the conformal Hubble rate, and $\Omega_{\hi}(z)$ is the comoving HI density in units of the critical density today, which is currently poorly constrained by observations and is modelled by simulations. We use the fit 
in  \cite{Santos:2017qgq}:
\be 
\bar{T}_{\mathrm{HI}}(z) = 0.0 56 +0.23\,z -0.024\, z^{2} ~~ \mathrm{mK}. \label{e1.24}
\ee}

The temperature fractional perturbation is
\begin{equation}
\Delta_{\rm HI}(z,\bm{n}) = \frac{T_{\rm HI}(z,\bm{n}) - \bar {T}_{\rm HI} (z)}{\bar {T}_{\rm HI} (z)}\,.\label{e1.1_02}
\end{equation}
Using \eqref{tobs}, this leads to the following perturbative expansion (our convention is $X+X^{\tw}/2$).
\red{(A clear and concise derivation of the following expressions for $\Delta$ and $\Delta^{(2)}$ is given in Appendix A of \cite{DiDio:2018zmk}.)}
\begin{itemize}
\item
{\bf At first order} \cite{Hall:2012wd}:
\be \label{dt1}
\Delta \equiv \Delta^{(1)}_{\hi} = \Delta_{\rm N} + \Delta_{\rm D}\,,\quad  \Delta_{\rm N} = b_1\delta_{\rm m}-{1\over \H}\p_r(\v\cdot\n )\,,\quad \Delta_{\rm D}=A\,(\v\cdot\n ) \,.
\ee
Here   $r$ is the radial comoving distance and $\v=\bm{\nabla}V$ is the peculiar velocity.
$ \Delta_{\rm N} $ is the standard density + RSD term, which scales as $\delta_{\rm m}$. $\Delta_{\rm D}$ is the dominant relativistic correction, scaling as i\,$(\cH/k)\delta_{\rm m}$ in Fourier space. This Doppler term has coefficient
\be
A = b_{\rm e} - 2- \frac{\cH'}{\cH^{2}}=-{\ud \ln \big[(1+z) \bar{T}_{\hi} \big] \over \ud \ln (1+z)} \,,
 \label{e1.11}
\ee
where  the evolution bias is \cite{Fonseca:2015laa}
\be
 b_{\rm e} = -{\ud \ln \big[(1+z)^{-1}\H\, \bar{T}_{\hi} \big] \over \ud \ln (1+z)}\,.
\ee
We omit sub-leading relativistic corrections that scale as $(\H/k)^2\delta_{\rm m}$.

\item
{\bf At second order}
\cite{Maartens:2019yhx} (see also \cite{Umeh:2015gza,DiDio:2015bua,Umeh:2016thy,Clarkson:2018dwn,DiDio:2018zmk}):
\bea
\Delta^{\tw} &\equiv & \Delta^{(2)}_{\hi} = \Delta^{\tw}_{\rm N} + \Delta^{\tw}_{\rm D} \,,\\
\Delta^{\tw}_{\rm N}&=& b_1\delta_{\rm m}^{\tw}+b_2 \big(\delta_{\rm m}\big)^2+ b_{s^2}s^2 + {\rm RSD}^{\tw}\,,
\label{d2n} \\
\Delta^{\tw}_{\rm D} &=& A\, (\bm{v}^{(2)}\!\!\cdot\bm{n})+2\Big[ b_{1}\big(A+f) + \frac{b_1'}{\cH}\Big]\,(\bm{v}\cdot\bm{n})\,\delta_{\rm m} +\frac{1}{\cH}\big( 8-4A-3\Omega_{\rm m}   \big)(\bm{v}\cdot\bm{n})\,\partial_r(\bm{v}\cdot\bm{n})
\nonumber\\ \label{e1.2}
&&{}+\frac{2}{\cH^2}\big[(\bm{v}\cdot\bm{n}) \,\partial_r^2\Phi-\Phi\, \partial_r^2 (\bm{v}\cdot\bm{n}) \big]
 -\frac{2}{\cH}\,\partial_r (\bm{v}\cdot\bm{v})+2 \frac{{b_{1}}}{\cH}\,\Phi\, \partial_r\delta_{\rm m} \,.
\eea
In \eqref{d2n}, $b_{s^2}s^2$ is the tidal bias contribution and RSD$^{\tw}$ is the standard second-order RSD contribution (see \cite{Maartens:2019yhx} for details). The bias parameters are computed via a halo model (see  Appendix \ref{app1}) and are shown in Figure \ref{fig1}, together with the evolution bias.
In \eqref{e1.2}, we see the Doppler terms and the line-of-sight gradients that make up the dominant relativistic contribution.
 $\Phi$ is the gravitational potential and $\Omega_{\rm m} =\Omega_{{\rm m}0}(1+z)H_0^2/\cH^2$. We  neglect sub-dominant relativistic effects in $\Delta^{\tw}$ that scale as  $(\H/k)^2(\delta_{\rm m})^2$. 
%
\begin{figure}[h]
\centering
\includegraphics[width=.49\textwidth]{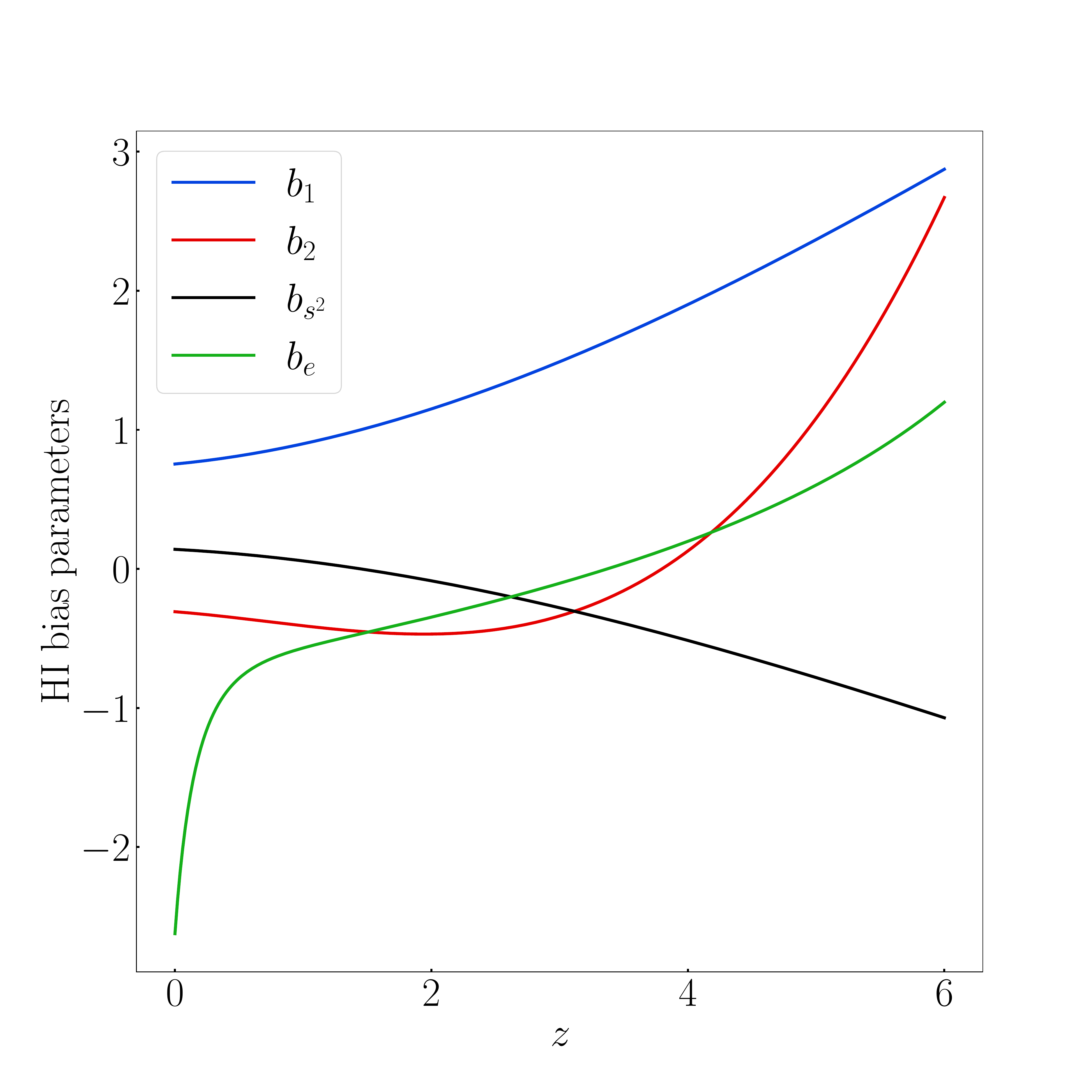}
\includegraphics[width=.49\textwidth]{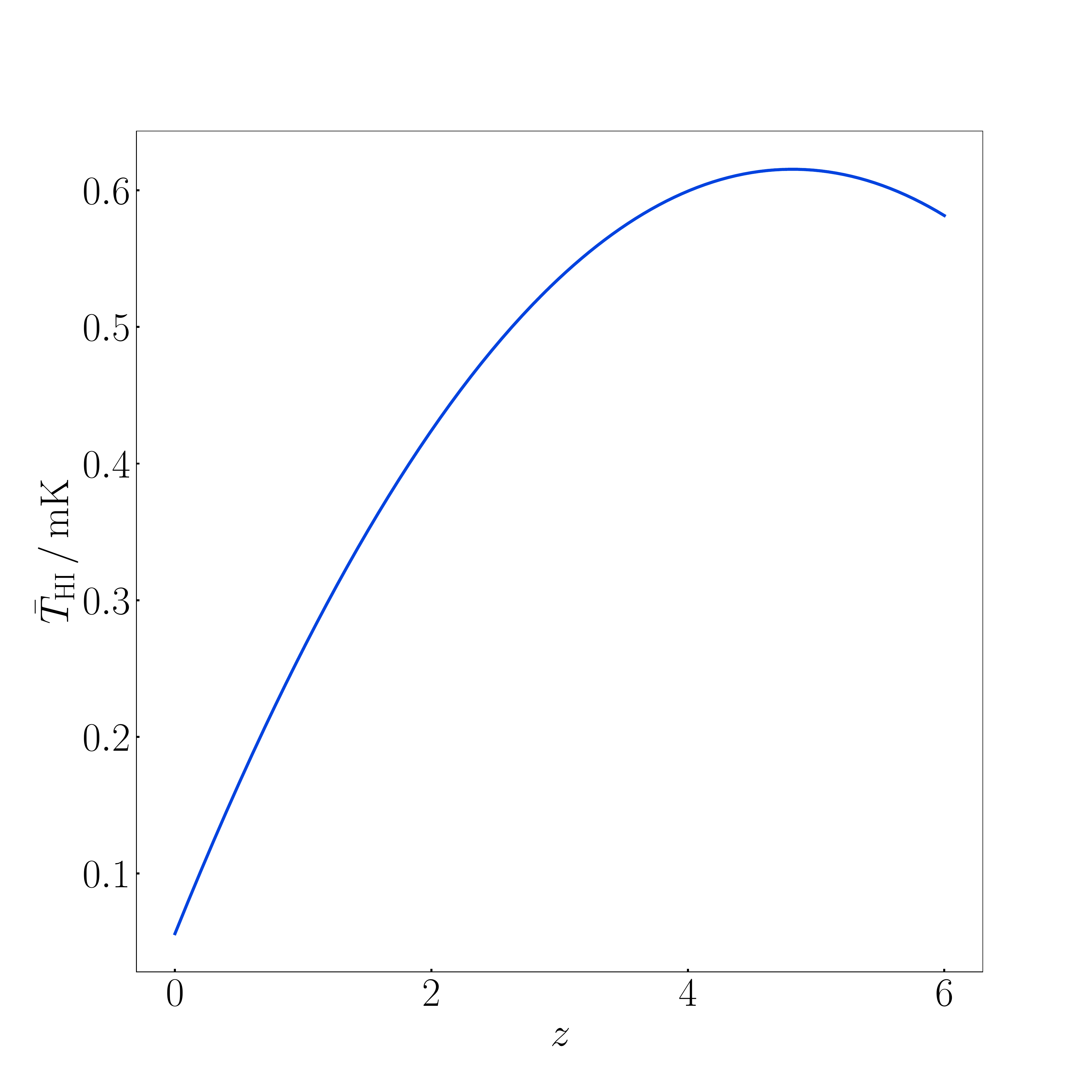}
\caption{HI clustering and evolution bias parameters (left) and background temperature (right).}\label{fig1}
\end{figure}

\item {\bf Lensing contribution:}\\
At first order, there is no lensing contribution to $\Delta$ \cite{Hall:2012wd}. The general case of galaxy number density contrast contains a lensing  contribution $2({\cal Q}-1)\kappa$ to $\Delta_g$, where $\kappa$ is the convergence \cite{Challinor:2011bk}.  For HI emitters in intensity mapping, the magnification bias satisfies
\be
{\cal Q} \equiv -{\p \ln \bar{N}_{\hi} \over \p \ln L}\bigg|_{\rm c} = 1\,, \label{magb}
\ee
where c indicates evaluation at the luminosity cut. 

At second order,  \eqref{tobs} shows that there is also no contribution to $\Delta^{\tw}$ from lensing convergence \cite{DiDio:2015bua,Jalivand:2018vfz}. We can recover this result from the full general expression for second-order number density contrast  \cite{Bertacca:2014dra, Bertacca:2014wga, Yoo:2014sfa, DiDio:2014lka, Bertacca:2014hwa}, by imposing \eqref{magb} together with the conditions \cite{DiDio:2015bua}
\be \label{magb2}
{\p^2 \ln \bar{N}_{\hi}  \over \p (\ln L)^2}\bigg|_{\rm c} = 0\,, \quad  {\p b_1 \over \p \ln L}\bigg|_{\rm c}=0\,.
\ee

There remains however a lensing deflection contribution $\nabla_{\perp a}\Delta\,\nabla_\perp^a\phi$ to  $\Delta^{\tw}$,  where $\nabla_{\perp a}$ is a screen-space gradient and $\phi$ is the lensing potential \cite{Umeh:2015gza,DiDio:2015bua,Jalivand:2018vfz}. In the bispectrum the contribution of this term is negligible for equal-redshift correlations \cite{DiDio:2015bua,Durrer:2020orn}. Since we only consider the bispectrum at equal redshifts, we can safely neglect this term. 
\end{itemize}}
\vspace{0.2cm}

In Fourier space, the  HI bispectrum  at  tree level  is defined by
\begin{equation}
\big\langle \Delta(z,\bm{k}_{1})\Delta(z,\bm{k}_{2})\Delta^{(2)}(z,\bm{k}_{3}) \big\rangle + \text{2 cp}=2 (2\pi)^3 B_{\rm HI}(z, \bm{k}_{1}, \bm{k}_{2}, \bm{k}_{3}) \delta^{\rm Dirac}\big(\bm{k}_{1}+ \bm{k}_{2}+ \bm{k}_{3} \big)\,, \label{e1.3}
\end{equation}
where cp denotes cyclic permutation. It follows that
\begin{equation}
B_{\rm HI}(z, \bm{k}_{1}, \bm{k}_{2}, \bm{k}_{3}) = \mathcal{K}^{(1)}(z, \bm{k}_{1})\mathcal{K}^{(1)}(z, \bm{k}_{2})\mathcal{K}^{(2)}(z, \bm{k}_{1}, \bm{k}_{2}, \bm{k}_{3})P_{\rm m} (z, k_{1})P_{\rm m} (z, k_{2}) + \text{2 cp}\;, \label{e1.5}
\end{equation} 
where $P_{\rm m} $ is the linear matter power spectrum (computed using CLASS \cite{Blas:2011rf}).  From now on, we often drop the $z$ dependence for brevity.  
The  bispectrum kernels are as follows.
\begin{itemize}
\item
{\bf Standard (Newtonian) kernels:}
\begin{align}
\mathcal{K}^{(1)}_{\rm N}(\bm{k}_{a}) &= b_{1}+f\mu_{a}^{2}\,,  \label{e1.7} \\ 
\mathcal{K}^{(2)}_{\rm N}(\bm{k}_{1}, \bm{k}_{2},{\bm{k}_3}) &= b_{1}F_{2}(\bm{k}_{1}, \bm{k}_{2}) + b_{2} + f\mu_{3}^{2}G_{2}(\bm{k}_{1}, \bm{k}_{2}) +{fZ_2}(\bm{k}_{1}, \bm{k}_{2})
+ b_{s^{2}}S_{2}(\bm{k}_{1}, \bm{k}_{2})\,, \label{e1.8}
\end{align}
where $f$ is the linear matter growth rate, $\mu_a=\hat{\k}_a\cdot\n$ and the standard $F_2,G_2,Z_2,S_2$ kernels are given in \cite{Maartens:2019yhx}.

\item
{\bf Leading-order relativistic kernels:} 
\begin{align}
\mathcal{K}^{(1)}_{\mathrm{D}}(\bm{k}_{a}) &= \mathrm{i}\,\cH f A\,\frac{\mu_{a}}{k_{a}}\,, \label{e1.9} \\
\mathcal{K}^{(2)}_{\mathrm{D}}(\bm{k}_{1},\bm{k}_{2},\bm{k}_{3}) &= \mathrm{i}\,\cH f \bigg\{
A\,\frac{\mu_{3}}{k_{3}}G_{2}(\bm{k}_{1},\bm{k}_{2})
+{\Big[b_1\big(A+f\big)+{b_1' \over \H} \Big]}\Big(\frac{\mu_{1}}{k_{1}} + \frac{\mu_{2}}{k_{2}}\Big)
\nonumber \\
&  -\frac{3}{2}\Omega_{\rm m} \Big(\mu_{1}^{3}\frac{k_{1}}{k_{2}^{2}} + \mu_{2}^{3}\frac{k_{2}}{k_{1}^{2}}\Big)
+{\Big[\frac{3}{2}\Omega_{\rm m} \big(1+f \big)+2f\big(A-2\big) \Big]}\mu_{1}\mu_{2}\Big(\frac{\mu_{1}}{k_{2}}+\frac{\mu_{2}}{k_{1}}\Big)
\nonumber \\
& +2f\,  {\hat{\bm{k}}_{1} \cdot \hat{\bm{k}}_2}\Big(\frac{\mu_{1}}{k_{1}} + \frac{\mu_{2}}{k_{2}}\Big) 
-\frac{3\Omega_{\rm m} b_1}{2f}\Big(\mu_{1}\frac{k_{1}}{k_{2}^{2}} + \mu_{2}\frac{k_{2}}{k_{1}^{2}}\Big)\!  \bigg\}\,,\label{e1.10}
\end{align}
where $A$ is given by \eqref{e1.11}. These follow from \eqref{dt1} and \eqref{e1.2}, and {agree with \cite{Maartens:2019yhx} when we impose \eqref{e1.11}, \eqref{magb} and \eqref{magb2}}.
\item
{\bf Complex bispectrum:}\\
From \eqref{e1.7}--\eqref{e1.10} we see that
$B_{\rm HI}$ is complex: {the imaginary part is given purely by local relativistic corrections, while at leading order, i.e. neglecting relativistic terms of $O(\H^2/k^2)$, the real part is given purely by the standard Newtonian bispectrum:}
\bea
\mathrm{Re}\big(B_{\rm HI}\big) &= &B_{\rm N} = \mathcal{K}^{(1)}_{\rm N}(\bm{k}_{1})\mathcal{K}^{(1)}_{\rm N}(\bm{k}_{2})\mathcal{K}^{(2)}_{\rm N}(\bm{k}_{1}, \bm{k}_{2}, \bm{k}_{3})P_{\rm m} (k_{1})P_{\rm m} (k_{2}) + \text{2 cp}, \label{e1.14}\\
{\mathrm{i \,Im}}\big(B_{\rm HI}\big) &=& B_{\rm D}= \bigg\{\bigg[\mathcal{K}^{(1)}_{\mathrm{N}}(\bm{k}_{1})\mathcal{K}^{(1)}_{\mathrm{D}}(\bm{k}_{2}) + \mathcal{K}^{(1)}_{\mathrm{D}}(\bm{k}_{1})\mathcal{K}^{(1)}_{\mathrm{N}}(\bm{k}_{2})\bigg]\mathcal{K}^{(2)}_{\mathrm{N}}(\bm{k}_{1},\bm{k}_{2},\bm{k}_{3}) 
\label{e1.15}\\ \nonumber 
&&{}~~~~~~~~~~~~~ +\mathcal{K}^{(1)}_{\mathrm{N}}(\bm{k}_{1})\mathcal{K}^{(1)}_{\mathrm{N}}(\bm{k}_{2})\mathcal{K}^{(2)}_{\mathrm{D}}(\bm{k}_{1},\bm{k}_{2},\bm{k}_{3})\bigg\}P_{\rm m} (k_{1})P_{\rm m} (k_{2})+\text{2 cp}.
\eea
It is apparent that {$B_{\rm N} \sim P_{\rm m} ^2$ while $B_{\rm D} \sim {\rm i}\, (\H/k) P_{\rm m} ^2$.}
Equation \eqref{e1.15} makes explicit the coupling of relativistic  and Newtonian terms in the bispectrum.
\end{itemize}


\section{Relativistic signal-to-noise ratio}

{The signal-to-noise ratio for a bispectrum in the Gaussian case is given by 
${\rm SNR}^2 = {B^2 / {\rm Var}(B)}$, where the variance includes cosmic variance as well as noise.
The relativistic part of the bispectrum at leading order is extracted as the imaginary part of the estimator for the bispectrum. For the variance, we use the fact that it is unaffected by relativistic corrections at leading order: ${\rm Var}(B_{\hi})= {\rm Var}(B_{\rm N})+O(\H^2/k^2)$, as shown in \cite{Maartens:2019yhx}.}

The bispectrum in each $z$-bin has 5 independent degrees of freedom, which we choose as the 3 sides $k_a$ of the closed triangle and 2 angles $(\theta_1=\cos^{-1}\mu_1,\varphi)$
that define the orientation of the triangle.
The relativistic SNR per $z$-bin is then given by  \cite{Maartens:2019yhx}
\begin{equation}
{{\rm SNR}(z)^2 =
\sum_{k_a,\mu_1,\varphi}\,\frac{B_{\rm D}(z, k_{a},  \mu_{1},\varphi) \,B_{\rm D}^*(z, k_{a},  \mu_{1},\varphi)}{{\rm Var} [{B_{\rm N}}(z, k_a,\mu_{1},\varphi)]}\,,} \label{e2.2_0}
\end{equation}
where we isolated the imaginary part, 
and the cumulative SNR is defined by
\begin{equation}
{{\rm SNR}(\leq z)^2 =
\sum_{z'}\,{\rm SNR}(z')^2\,.} \label{csnr}
\end{equation}
The total SNR is  ${\rm SNR}(\le z_{\rm max})$.

The variance is estimated as 
\begin{equation}
\!\!\!{{\rm Var} [{B_{\rm N}}(z, k_a,\mu_{1},\varphi)]}
=\frac{s_B\,\pi\, k_{\rm f}(z)^3}{k_1k_2k_3 (\Delta k)^3}\,\frac{{4\pi}}{ \Delta \mu_1 \Delta \varphi} \,
\tilde{P}_{\rm HI}(z,k_{1},\mu_{1})\,\tilde{P}_{\rm HI}(z,k_{2},\mu_{2}) \,\tilde{P}_{\rm HI}(z,k_{3},\mu_{3})\,.
\label{e2.7}
\end{equation}
Here
\begin{equation}
\tilde{P}_{\rm HI}(z, k, \mu) = P_{\rm HI}(z, k, \mu) + P_{\rm noise}(z,k,\mu)\,, \label{e2.4}
\end{equation}
where the HI power spectrum is Newtonian at leading order: $P_{\rm HI}= P_{\rm N} +O(\H^2/k^2)$, with
\begin{equation}
P_{\rm N}(z, k, \mu) = \big[b_{1}(z)+f(z)\mu^{2}\big]^{2}P_{\rm m}(z,k)\,.\label{e2.5_1}
\end{equation}
In \eqref{e2.7}, the multiplicity constant $s_B=6, 2, 1$ for equilateral, isosceles, and non-isosceles triangles,  and $\Delta k,\Delta \mu_1,  \Delta \varphi$ are bin widths. We follow \citep{Karagiannis:2018jdt,Maartens:2019yhx} and choose the step lengths as
\begin{equation}
\Delta z = 0.1\,, \quad \Delta \mu = 0.04\,, \quad {\Delta\varphi=\pi/25}\,, \quad \Delta k(z) = k_{\mathrm{f}}(z)\,.\label{e4.1} 
\end{equation} 
The fundamental mode $k_{\mathrm{f}}$ is an estimate of  the minimum wavenumber included in the survey, and is determined in each redshift bin  by the comoving volume of the bin centred at $z$:
\begin{equation}
k_{\mathrm{f}}(z) = \frac{2\pi}{V(z)^{1/3}} \quad \mbox{with} \quad V(z) = \frac{4\pi}{3}f_{\mathrm{sky}} \big[r(z+{\Delta z}/{2})^{3} - r(z-{\Delta z}/{2})^{3} \big] \,.\label{e2.8}
\end{equation}
Here $f_{\mathrm{sky}}=\Omega_{\rm sky}/4\pi$ is the sky fraction covered by the survey.

\subsection{Nonlinearity}

The tree-level bispectrum is a perturbative model. In order to avoid scales where dark matter clustering becomes nonperturbative, we impose a maximum scale \cite{Maartens:2019yhx}
\be \label{kmax}
k_{\mathrm{max}}(z) = 0.1h\,(1+z)^{2/(2+n_s)}~{\rm Mpc}^{-1}\,.
\ee
However, applying this  cut-off still includes nonperturbative RSD effects, i.e. the fingers-of-god damping. In order to take account of this, we follow \cite{Karagiannis:2018jdt,Yankelevich:2018uaz,Maartens:2019yhx} and use the following \red{phenomenological model of the damping of redshift-space clustering due to nonlinear velocities}:
\begin{align}
 P_{\rm HI}(k,\mu,z) & ~\to~   \exp\Big[-\frac{1}{2}{k^2\mu^2\,{\sigma_{\red{P}}(z)}^{2}}\Big]\,P_{\rm HI}(k,\mu,z)
 \,, \label{e2.10}\\
 B_{\rm HI}(k_a,\mu_a,z) & ~\to~  \exp\Big[-\frac{1}{2}{\left( k_{1}^{2}\mu_{1}^{2}+k_{2}^{2}\mu_{2}^{2}+k_{3}^{2}\mu_{3}^{2}\right){\sigma_{\red{B}}(z)}^{2}}\Big]\,B_{\rm HI}(k_a,\mu_a,z)
 \,. \label{e2.11}
\end{align}
\red{Damping is effective for $k\gtrsim \sqrt{2}/\sigma_P$ for the power spectrum and  $k\gtrsim \sqrt{2}/\sigma_B$ for the bispectrum.
In this model, the damping parameters $\sigma_P$ and $\sigma_B$ are  to be determined by data or simulations.
We use a model for the power spectrum damping parameter based on HI simulations \cite{Sarkar:2019nak}:}
\be
\sigma_{\red{P}}(z) = 11h\big(1+z\big)^{-1.9}\exp{\big[-(z/11)^{2}\big]}~~h^{-1}\,\mathrm{Mpc}\,. \label{e2.12}  \\
\ee
\red{We are not aware of any simulation-based expression for $\sigma_B$, and we therefore take $\sigma_B=\sigma_P$ as a first approximation. The
effect of increasing $\sigma_B$ is discussed in Section \ref{sec5}.}

A further source of nonlinearity on weakly nonlinear scales that contributes to the bispectrum variance  is the non-Gaussian effect due to mode coupling. {We follow \cite{Karagiannis:2018jdt,Maartens:2019yhx} and  take account of this using the simple model developed in \cite{Chan:2016ehg}, which modifies the variance as follows: 
\be
{\rm Var} [{B_{\rm N}}(\k_a)]  \to  {\rm Var}  [{B_{\rm N}}(\k_a)] \left[1+  
{\delta\tilde{P}_{\rm N}(\k_1)\over \tilde{P}_{\rm N}(\k_1)} + {\delta\tilde{P}_{\rm N}(\k_2)\over \tilde{P}_{\rm N}(\k_2)} + {\delta\tilde{P}_{\rm N}(\k_3)\over \tilde{P}_{\rm N}(\k_3)}\right]  .
 \label{e2.14}
 \ee
 Here the redshift dependence  has been dropped for brevity, and
 \bea
\delta\tilde{P}_{\rm N}(\k) & =& \tilde{P}_{\rm N}^{\rm nl}(\k) - \tilde{P}_{\rm N}(\k)\,, 
\label{e2.15}\\
\tilde{P}^{\rm nl}_{\rm N}(\k) &=& \big(b_{1}+f\mu^2\big)^2P^{\rm nl}_{\rm m}(k)+P_{\rm noise}(\k)\,,\label{e2.15_1}
\eea
where \red{$\tilde{P}_{\rm N}={P}_{\rm N}+P_{\rm noise}$} and
 $P^{\rm nl}_{\rm m}$ is the nonlinear matter power spectrum computed in CLASS with a modified Halofit emulator.} 

\subsection{{Effects of foregrounds}}

{Foreground contamination is the major systematic confronting 21cm intensity mapping. 
Cleaning techniques are very efficient at recovering the cosmological signal in regions of $(k_\|,k_\perp)$ space (see e.g. \cite{Pober:2013jna, Bull:2014rha, Alonso:2014dhk, Pober:2014lva, Wolz:2015sqa, Santos:2015gra, Shaw:2014khi, Obuljen:2017jiy, Ansari:2018ury, Witzemann:2018cdx, Bacon:2018dui, Asorey:2020mxs, Cunnington:2020mnn}). A realistic model of bispectrum measurements should include modelling of foreground removal. However, our focus is on the relativistic signal in the bispectrum and so we take a simpler approach -- by excising the regions of $(k_\|,k_\perp)$ space where foreground cleaning does not recover  the signal efficiently.}

HI intensity mapping 
surveys are planned for next-generation radio dish arrays, including
MeerKAT\footnote{www.sarao.ac.za/science/meerkat/}, 
SKA1-MID\footnote{{www.skatelescope.org/}} and
HIRAX\footnote{hirax.ukzn.ac.za/}. We will also consider surveys with
PUMA\footnote{{www.puma.bnl.gov/}}  in its initial phase (Petite) (note that PUMA is currently still a proposal).
 Intensity mapping surveys
can be done in 2 survey modes:
\begin{itemize}
\item Single-dish (SD) mode: auto-correlation signals from single dishes are added.
\item Interferometer (IF) mode: cross-correlation signals from array elements are combined. 
\end{itemize}  
In both SD- and IF-mode surveys, foreground cleaning effectively removes large-scale radial modes {because of the smoothness in freqency of the main foreground emissions}. \red{In order to model the effects of foreground cleaning, we remove radial modes $k_\| \equiv \mu k$ that are smaller than a critical scale $k_{\| \rm fg}$  \cite{Karagiannis:2018jdt}.
We impose this via an exponential damping factor \cite{Bull:2014rha}:  
\begin{equation}\label{e2.21}
D_{\rm fg}(k,\mu)= 1-\exp\left[-\left({\mu k \over k_{\| \rm fg}}\right)^2\right]\qquad \mbox{for SD and IF survey modes.}
\end{equation}
Then 
\begin{align}
P_{\rm HI}(k,\mu,z)  &~\to~ D_{\rm fg} (k,\mu)\,P_{\rm HI}(k,\mu,z) \,,\\
B_{\rm HI}(k_a,\mu_a,z) & ~\to~ D_{\rm fg} (k_1,\mu_1)\,D_{\rm fg} (k_2,\mu_2)\,D_{\rm fg} (k_3,\mu_3)\,B_{\rm HI}(k_a,\mu_a,z)\,.
\end{align}
The} value of $k_{\| \rm fg}$ can be reduced by techniques which reconstruct long modes from the information in measured short modes. This technique has been applied to HI intensity mapping by \cite{Zhu:2016esh,Modi:2019hnu}. Taking into account reconstruction, we choose 
\be \label{kfgpar}
k_{\| \rm fg} = 0.01 \,h\,{\rm Mpc}^{-1},
\ee
{but we will also consider an optimistic case, k$_{\| \rm fg} = 0.005 \,h\,{\rm Mpc}^{-1}$, that anticipates further developments of the reconstruction technique.}

IF-mode surveys lose additional signal due to the fact that {interferometers are chromatic, i.e., a fixed physical baseline length probes different angular scales at different frequencies.} This causes smooth foregrounds to leak into high-$k_{\perp}$ modes \cite{Obuljen:2017jiy,Alonso:2017dgh,Ansari:2018ury}. The signal loss may be accounted for by excluding
the region known as the foreground wedge \cite{Pober:2014lva, Pober:2013jna}: 
\begin{equation} \label{e2.19}
\big|k_{\parallel}\big| > k_{\rm wedge}(z)\,k_{\perp} \qquad \mbox{for IF survey mode,}
\end{equation} 
where $k_\perp=\sqrt{1-\mu^2}\,k$ and  $k_{\rm wedge}$ is modelled as
\begin{equation}
k_{\rm wedge}(z) = r(z)\cH(z) \sin{\big[0.61N_{\mathrm{w}}\,{\theta_{\rm b}(z)}\big]} \, . \label{e2.20} 
\end{equation}
{The beam (defining the IF field of view) is given for a dish array by
\be\label{thetab}
\theta_{\rm b}(z) = 1.22 \,{\lambda(z) \over D_{\rm d}}\,,
\ee
where $\lambda(z)=\lambda_{21}(1+z)$ and $D_{\rm d}$ is the dish diameter.
$N_{\mathrm{w}}$ is the number of primary beams away from the beam centre that contaminate the signal. The wedge effect is a technical problem that can be mitigated (and  in principle removed) by calibration of baselines \cite{Ghosh:2017woo}.}
Following  \cite{Karagiannis:2019jjx}, we take
\be
N_{\mathrm{w}}= 0, 1, 3, 
\ee
where 0 is the most optimistic possibility (wedge removed by calibration)  and 3 is a pessimistic case. 

\subsection{{Maximum and minimum scales probed}}

{The nonlinearity limit $k\leq k_{\rm max}(z)$ given in \eqref{kmax}, defines a minimum wavelength $2\pi/ k_{\rm max}(z)$ that is independent of surveys and is determined purely by dark matter clustering. It applies to both SD- and IF-mode surveys.
However, in the case
of IF-mode surveys, there is also a lower limit on angular scales, i.e. an upper limit on transverse wavenumbers, via $k_\perp=2\pi/(r\theta)$. This arises because the maximum baseline $D_{\rm max}$ determines the  angular resolution  and the upper limit can be roughly estimated by a cut-off  \cite{Bull:2014rha,Alonso:2017dgh,Karagiannis:2019jjx,Durrer:2020orn}
\be \label{kifmax}
k_{\perp \rm max}^{\rm IF}(z) \approx {2\pi D_{\rm max}\over r(z)\,\lambda(z)}\,.
\ee
In our computations, we do not use this cut-off since it is effectively imposed by the baseline density factor (see Section \ref{inoise}).
\begin{figure}[h]
\centering
\includegraphics[width=.49\textwidth]{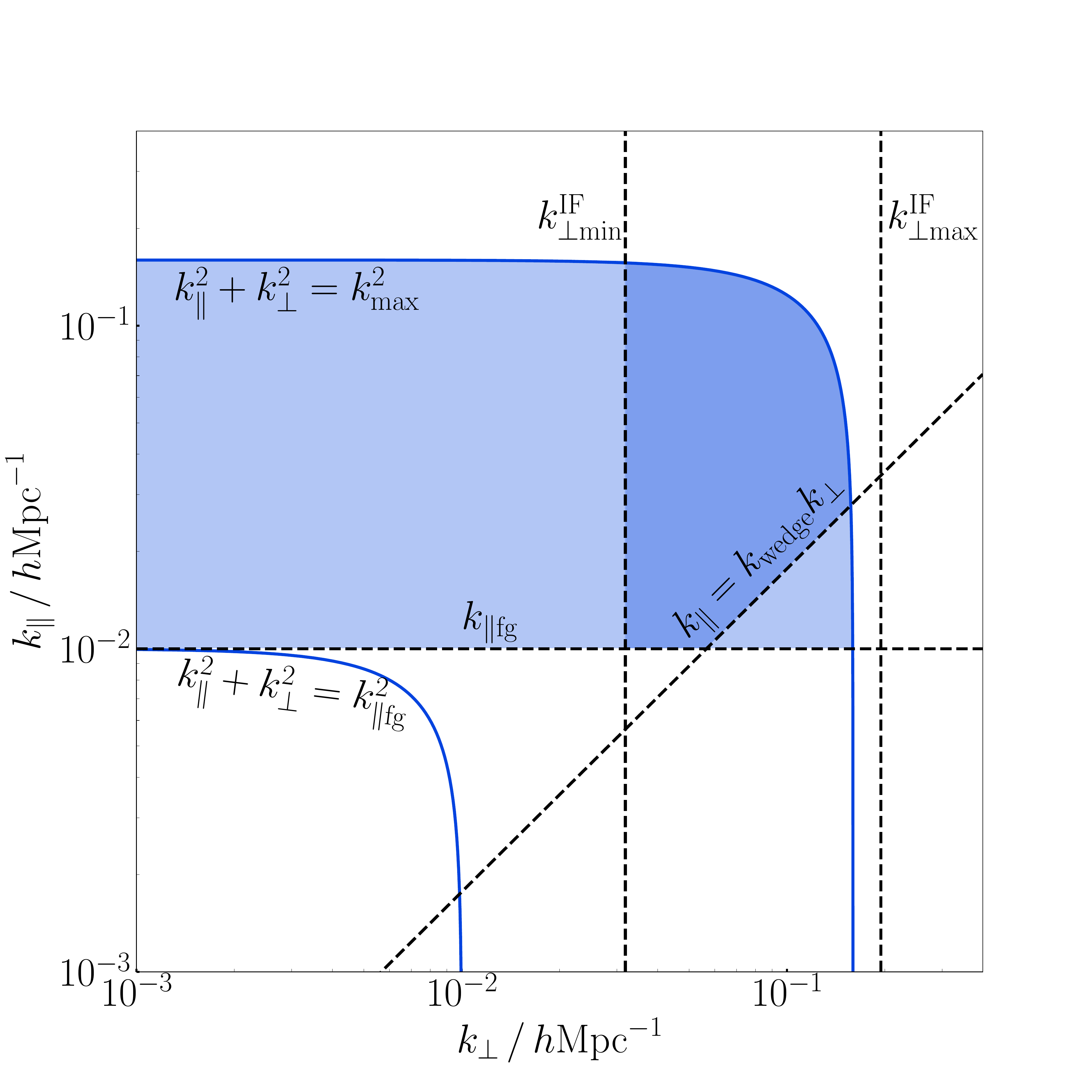}
\caption{{Schematic of important scales in the $k_\|,k_\perp$-plane at fixed $z$. Shading indicates the included region for SD mode (light) and IF mode (dark).}}\label{kplane}
\end{figure}

In principle, the maximum wavelength probed by HI intensity surveys at redshift $z$ is $2\pi/k_{\rm f}(z)$,  
where the fundamental mode $k_{\rm f}$ is determined by the comoving volume of the redshift bin, via \eqref{e2.8}. 
However, foreground cleaning imposes on both SD- and IF-mode surveys the  limiting minimum radial wavenumber $k_{\|\rm fg}$, given by \eqref{kfgpar}. Since $k^2=k_\|^2+k_\perp^2$, this means that  
\be \label{kmin}
k>k_{\|\rm fg} \quad \mbox{which implies}\quad k >  k_{\rm min}= {\rm max}\,\big\{ k_{\rm f}(z), k_{\|\rm fg} \big\}\,.
\ee
In IF-mode surveys there is a further minimum wavenumber, corresponding to the maximum angular scale that can be probed. This limit depends on the minimum baseline and is given by
 \cite{Bull:2014rha,Alonso:2017dgh,Karagiannis:2019jjx,Durrer:2020orn}:
\be \label{kifmin}
k_{\perp \rm min}^{\rm IF}(z) ={2\pi \over r(z)\,\theta_{\rm b}(z)}\,.
\ee
Since $k>k_\perp$, we have
\be \label{kminif}
k > k_{\rm min}=   {\rm max}\,\big\{ k_{\perp \rm min}^{\rm IF}(z), k_{\rm f}(z), k_{\|\rm fg} \big\} \quad \mbox{for IF surveys}.
\ee
The scales from foreground cleaning and the maximum and minimum scales are shown schematically in Figure \ref{kplane}. This does not include the effects of the beam in SD mode; 
see Figure \ref{pnoise} below.}

\subsection{Instrumental noise}\label{inoise}

{In HI intensity mapping for the scales and redshifts we consider, the shot noise is much smaller than the instrumental noise for next-generation surveys\footnote{{For the futuristic PUMA (Full) survey, the noise is low enough to be comparable to the shot noise \cite{Chen:2018qiu}.}} and can be safely neglected \cite{Castorina:2016bfm,Villaescusa-Navarro:2018vsg}. 
The noise power spectrum for the fractional temperature perturbation is then determined by instrumental noise and is given by \cite{Bull:2014rha,Alonso:2017dgh}:\footnote{It is also possible to put the beam factor $\beta$ in the signal, rather than in the noise; see e.g.  \cite{Bernal:2019jdo} for a discussion.}
\begin{equation}
{P}_{\mathrm{noise}}(z) = {2\pi f_{\mathrm{sky}}\over \nu_{21} t_{\mathrm{tot}}}\,\frac{(1+z) r(z)^{2} }{\H(z) }\,\left[\frac{T_{\mathrm{sys}}(z)}{\bar{T}_{\mathrm{HI}}(z)}\right]^2\, {\alpha(z,k_\perp) \over \beta(z,k_\perp)^2}
~~~ h^{-3}\mathrm{Mpc}^{3} \,. 
\label{e2.24}
\end{equation}
Here $t_\mathrm{tot}$ is the total observing time, and  the system temperature may be modelled as \citep{Ansari:2018ury}:
\begin{equation}
T_{\mathrm{sys}}(z) = T_{\rm d}(z)+T_{\rm sky}(z) =T_{\rm d}(z) + 2.7 + 25\bigg[\frac{400\,\mathrm{MHz}}{\nu_{21}} (1+z)\bigg]^{2.75} ~ \mathrm{K}, \label{e2.25}
\end{equation} 
where $T_{\rm d}$ is the dish receiver temperature. (We consider only dish arrays.)

In \eqref{e2.24}, the dish density factor $\alpha$ and the effective beam $\beta$ depend on the survey mode, as follows \cite{Bull:2014rha, Obuljen:2017jiy, Ansari:2018ury, Jalilvand:2019bhk, Durrer:2020orn}:
\bea
\alpha_{\rm SD} &=& {1\over N_{\rm d}}\,,\qquad\qquad\qquad\qquad  \beta_{\rm SD}= \exp\left[-\frac{k_\perp^2 r(z)^2\theta_{\rm b}(z)^2}{16\ln 2} \right], \label{sdn} \\
\alpha_{\rm IF} &=&\left[ {\lambda(z)^2\over A_{\rm e}}\right]^2 {1\over n_{\rm b}(z,k_\perp)}
\,,\quad~  \beta_{\rm IF}= \theta_{\rm b}(z)\,. \label{ifn}
\eea
Here $N_{\rm d}$ is the number of dishes  and $A_{\rm e}$ is the effective beam area:
\be \label{aeff}
A_{\rm e} = 0.7 A_{\rm d}\,, \quad A_{\rm d}={\pi \over 4}D_{\rm d}^2\,.
\ee
The dimensionless $n_{\rm b}$ is the baseline density in the image plane (assuming azimuthal symmetry), which is determined by the array distribution. {The total number of baselines is $\int \ud^2{\bm u}\,n_{\rm b}(u) = N_{\rm d}(N_{\rm d}-1)/2.$}
A physical baseline length $L$ is related to an image-plane scale $u$ as
\be
L = u\lambda = {k_\perp r \over 2\pi}\,\lambda\,.
\ee
Then the image-plane and physical distributions of the array are related by  \cite{Ansari:2018ury}
\be \label{nbphy}
n_{\rm b}(z,u) = \lambda(z)^2\, n_{\rm b}^{\rm phys}(L)\,.
\ee }


\subsection{Future HI intensity mapping surveys}

{We consider surveys proposed for the following dish arrays:}
\begin{itemize}
\item 
SD mode:  MeerKAT, 
SKA1-MID.
\item
IF\ mode: HIRAX,
PUMA (Petite). 
\end{itemize}
The survey specifications are given in Table~\ref{tab1}, {based on  \cite{Santos:2017qgq, Bacon:2018dui, Bandura:2019uvb,Karagiannis:2019jjx,Castorina:2020zhz}}. {Limiting wavenumbers for large wavelengths  are shown for these surveys in Figure \ref{figmin}: the
fundamental wavenumber \eqref{e2.8}, the minimum radial wavenumber \eqref{kfgpar} from foreground cleaning and the IF-mode minimum wavenumber \eqref{kifmin}.}

\begin{table}[h]
\centering
\caption{\label{tab1} {HI intensity mapping survey specifications. 
(For *, see Appendix \ref{app2}.)}} 
\vspace*{0.2cm}
\begin{tabular}{|lccccccc|} \hline
Survey & Redshift range & $f_{\rm sky}$ & $t_{\mathrm{tot}}$  & $T_{\mathrm{d}}$ & $D_{\rm d}$ & $N_{\rm d}$ & {$D_{\rm max}$} \\
&  & & [$10^{3}$ hr] & [K]  & [m] & & {[m]} \\
\hline\hline 
MeerKAT L Band   & 0.10--0.58 & 0.10 & 4  & * & 13.5 & 64 & -- \\
MeerKAT UHF Band & 0.40--1.45 & 0.10 & 4  & * & 13.5 & 64  & --  \\
SKA1-MID Band 1  & 0.35--3.05 & 0.48 & 10 & * & 15.0 & 197  & --  \\
SKA1-MID Band 2  & 0.10--0.49 & 0.48 & 10 & * & 15.0 & 197  & --  \\
HIRAX            & 0.75--2.00 & 0.36 & 10 & 50 & 6.0  & 1024  & {270}\\
PUMA (Petite)    & 2.00--6.00 & 0.50 & 40 & 50 & 6.0  & 5000 & {600} \\
\hline
\end{tabular}
\end{table}

\vspace*{-0.5cm}
\begin{table}[h]
\centering
\caption{\label{tab2} {Parameters in \eqref{e3.4}} (from \cite{Ansari:2018ury}).} 
\vspace*{0.2cm}
\begin{tabular}{|lcccccl|} \hline
Survey & $a$ & $b$ & $c$  & $d$ & $e$ & $N_{\rm s}$  \\ \hline\hline 
HIRAX & 0.4847 & $-0.3300$ & 1.3157 & 1.5974 & 6.8390 & 32  \\
PUMA (Petite) & 0.5698 & $-0.5274$ & 0.8358 & 1.6635 & 7.3177 & {100} \\
\hline
\end{tabular}
\end{table}

For the system temperature, we use the results of measurements and simulations for MeerKAT and SKA1-MID, given  in Appendix \ref{app2}. For HIRAX and PUMA, we use the fit \eqref{e2.25}.

In IF mode, HIRAX is assumed to be a square-packed array, while PUMA is taken as hexagonal-packed in a circular area, {with 50\% fill factor}. We follow \cite{Karagiannis:2019jjx} and use the fitting formula from \cite{Ansari:2018ury} for the baseline density of such arrays:
\begin{equation}
{n_{\rm b}^{\rm phys}(L) = \left({N_{\rm s} \over D_{\rm d}}\right)^2\,\frac{a+b\big(L/L_{\rm s}\big)}{1+c\big(L/L_{\rm s}\big)^{d}}\,\exp{\big[-(L/L_{\rm s})^{e}\big]}\,, } \label{e3.4}
\end{equation}
where  $L_{\rm s} =N_{\rm s}D_{\rm d}$ and $N_{\rm s}^2=N_{\rm d}$.
The  parameters in \eqref{e3.4} are given in Table \ref{tab2}, {and $n_{\rm b}^{\rm phys}(L)$ is shown in Figure \ref{nbphys}.
\begin{figure}
\centering
\includegraphics[width=.49\textwidth]{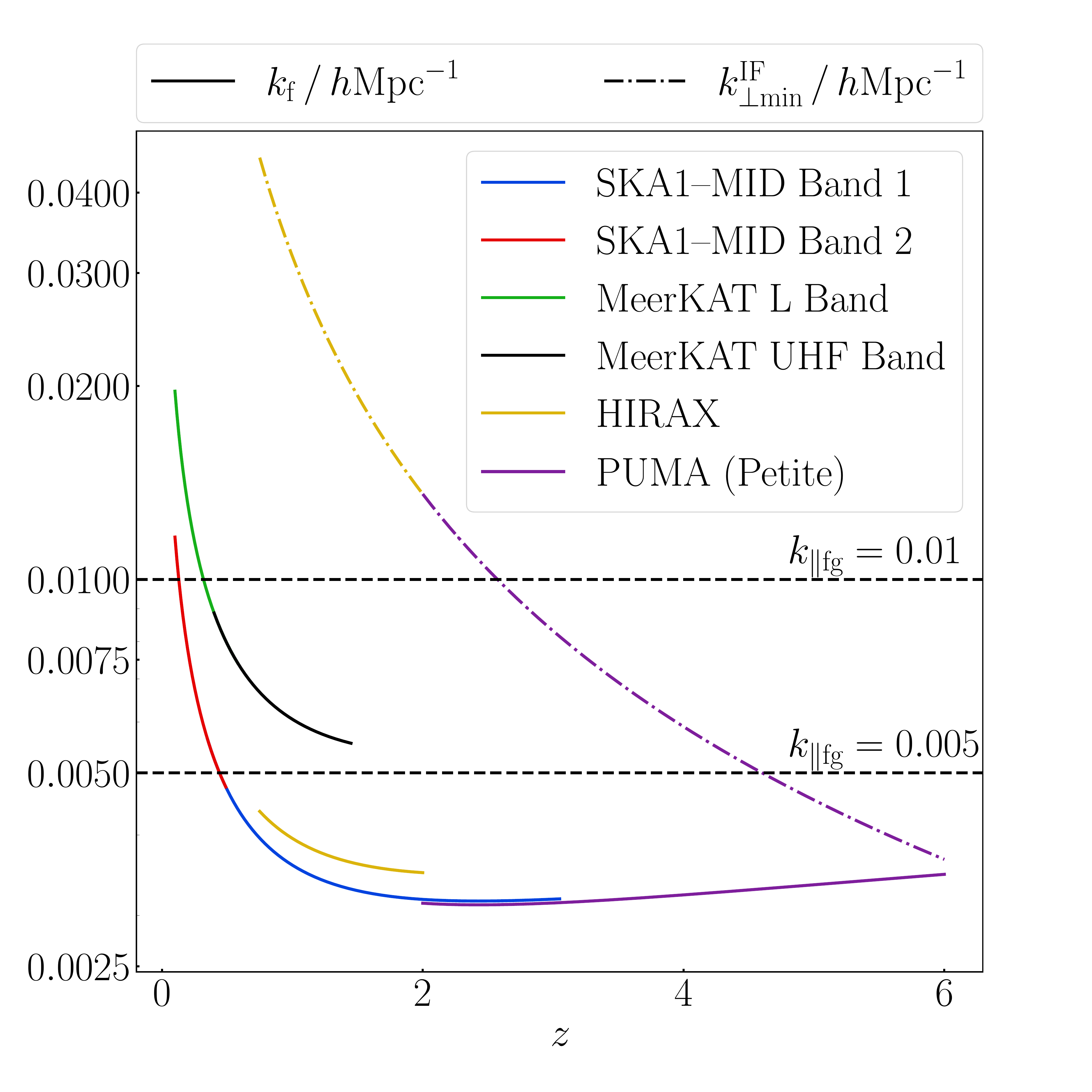}
\caption{{Minimum wavenumbers for the surveys, where $k$ is subject to \eqref{kmin} (SD) or \eqref{kminif} (IF).}}\label{figmin}
\end{figure}

\begin{figure}[h]
\centering
\includegraphics[width=.49\textwidth]{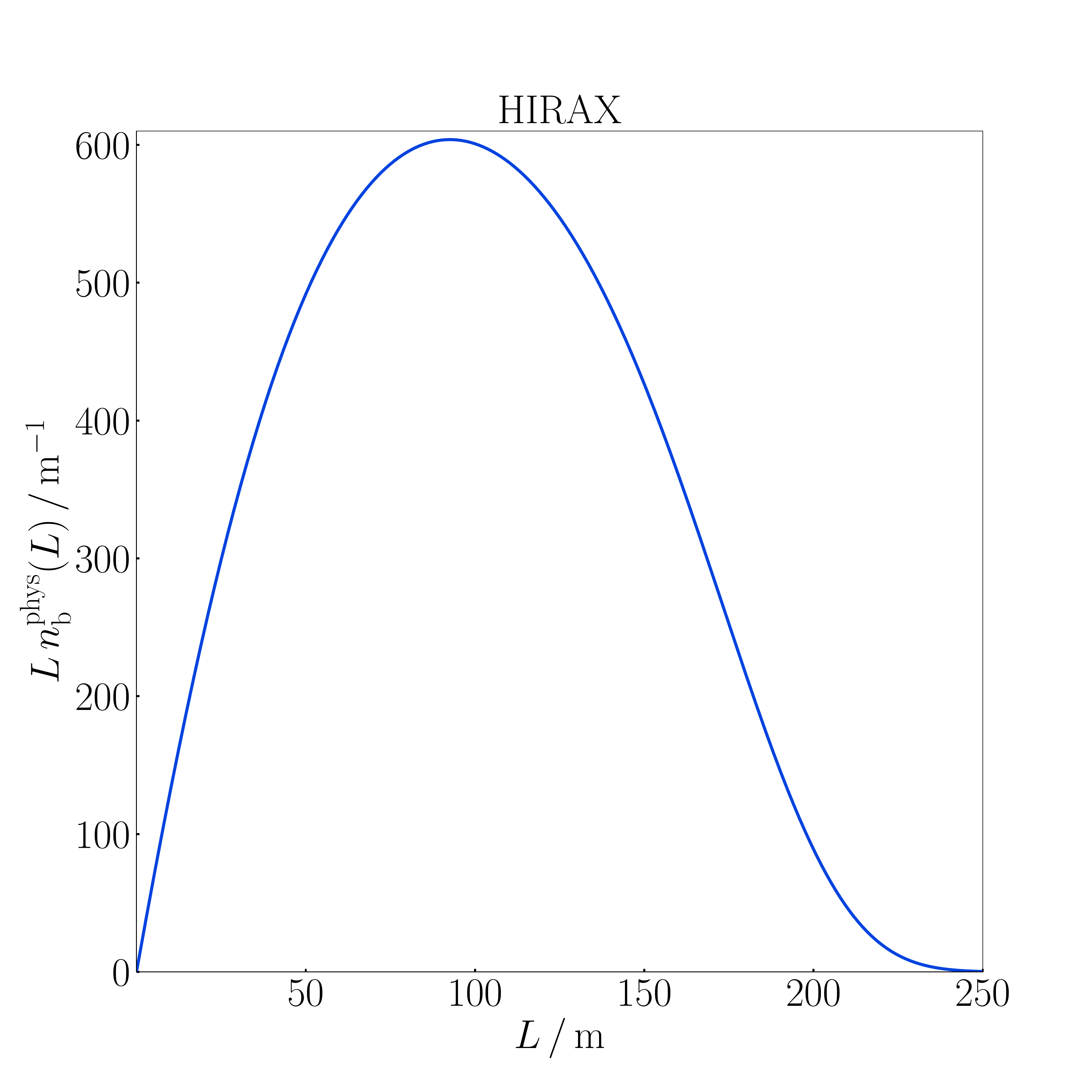}
\includegraphics[width=.49\textwidth]{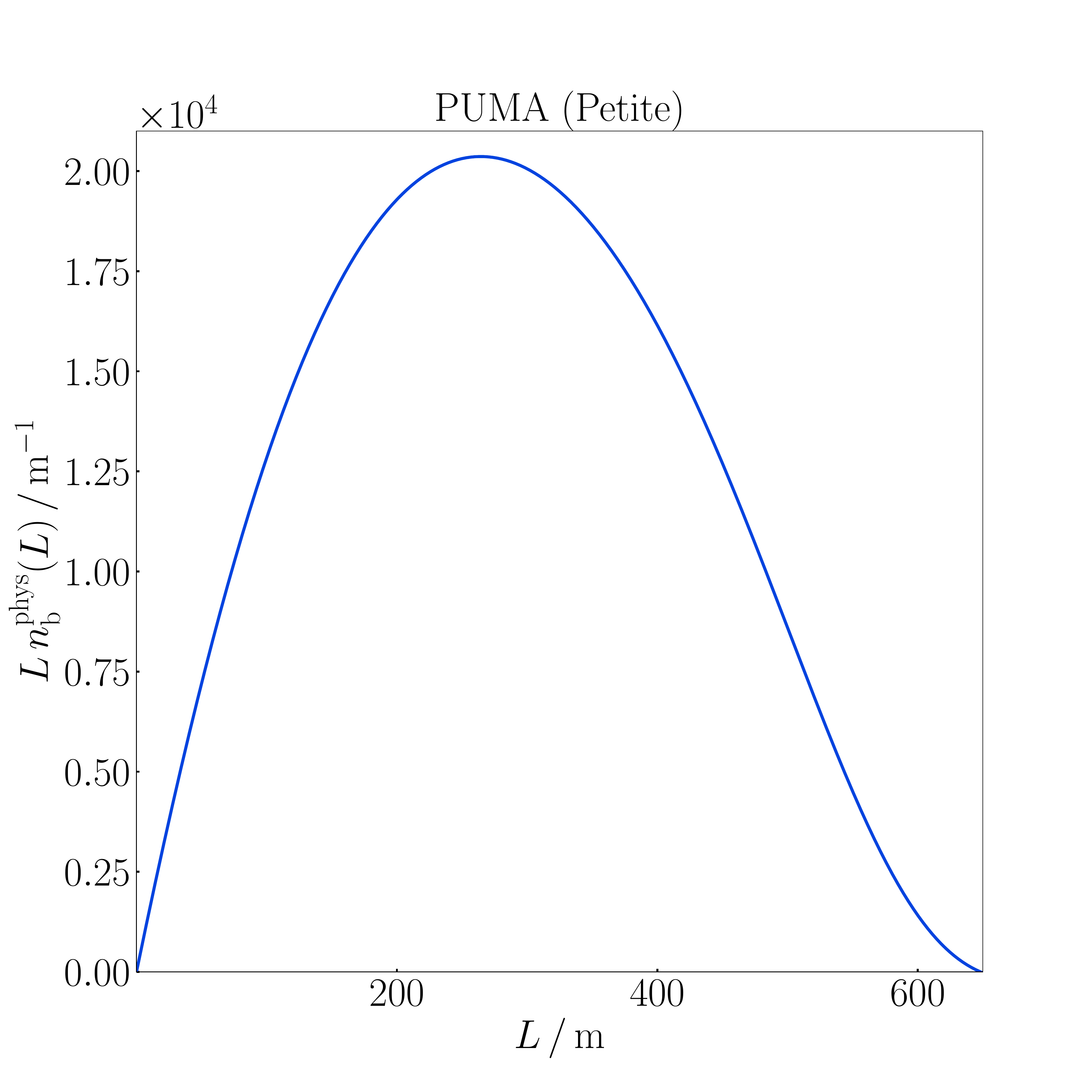}
\vspace*{-0.5cm}
\caption{Physical baseline density models for HIRAX (left) and PUMA (right).}\label{nbphys}
\end{figure}
\begin{figure}[!h]
\centering
\includegraphics[width=.49\textwidth]{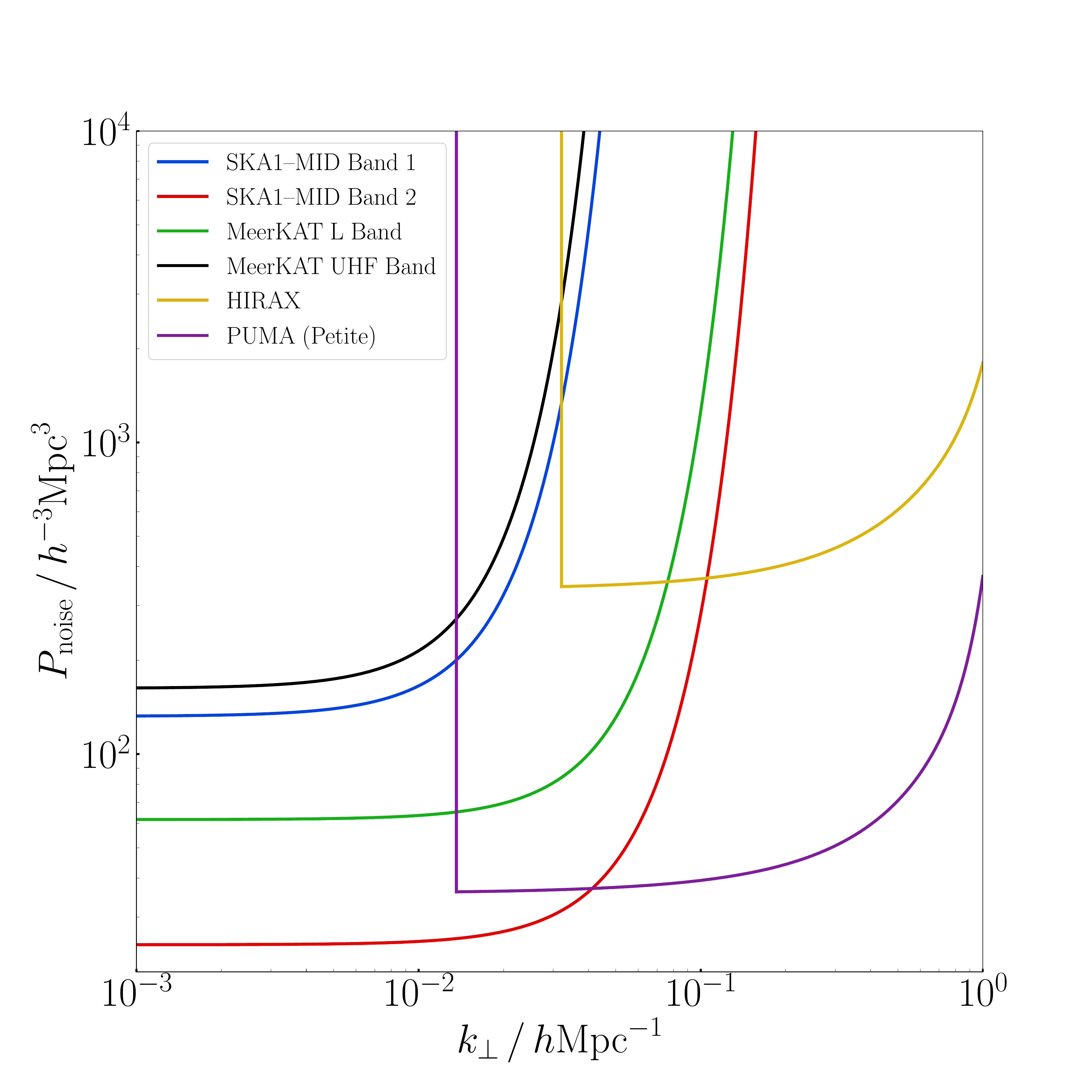}
\vspace*{-0.5cm}
\caption{{Noise power spectra of the SD-mode surveys (at $z=0.4$ for low-$z_{\rm max}$ bands and $z=1$ for high-$z_{\rm max}$ bands) and IF-mode surveys (HIRAX at $z=1$, PUMA at $z=2$).}}\label{pnoise}
\end{figure}

{In the case of PUMA, we take account of the 50\% fill factor as follows. We use double the number of dishes to define $N_{\rm s}$ for the computation of \eqref{e3.4}, i.e. $N_{\rm s}^2= 2\times 5000$. Then  we  remove half of the dishes without changing the baseline, i.e. without changing the shape and ground-area of the array.

The noise power spectra of the surveys at fixed redshift are displayed in Figure \ref{pnoise}. For SD-mode surveys, the poor angular resolution is reflected in the blow-up of noise due to the beam in \eqref{sdn}.
The minimum transverse scale for IF-mode surveys  is shown as a sharp cut-off, as in  \eqref{kifmin}, with effectively infinite noise.
 Figure \ref{pnoise} also shows the smooth blow-up of IF-mode noise on small transverse scales, which results  from the fact that 
$n_{\rm b}\to 0$ as the baseline approaches its maximum $D_{\rm max}$ (see Figure \ref{nbphys}). The unbounded increase of noise kills the signal, corresponding to the approximate cut-off scale \eqref{kifmax}.}


\section{Forecasts for the relativistic signal-to-noise}

\subsection{Single-dish mode}

The  SNR of the relativistic part of the bispectrum is shown in Figure~\ref{fig3}, per $z$-bin and cumulative.
\begin{figure}[h]
\centering
\includegraphics[width=.49\textwidth]{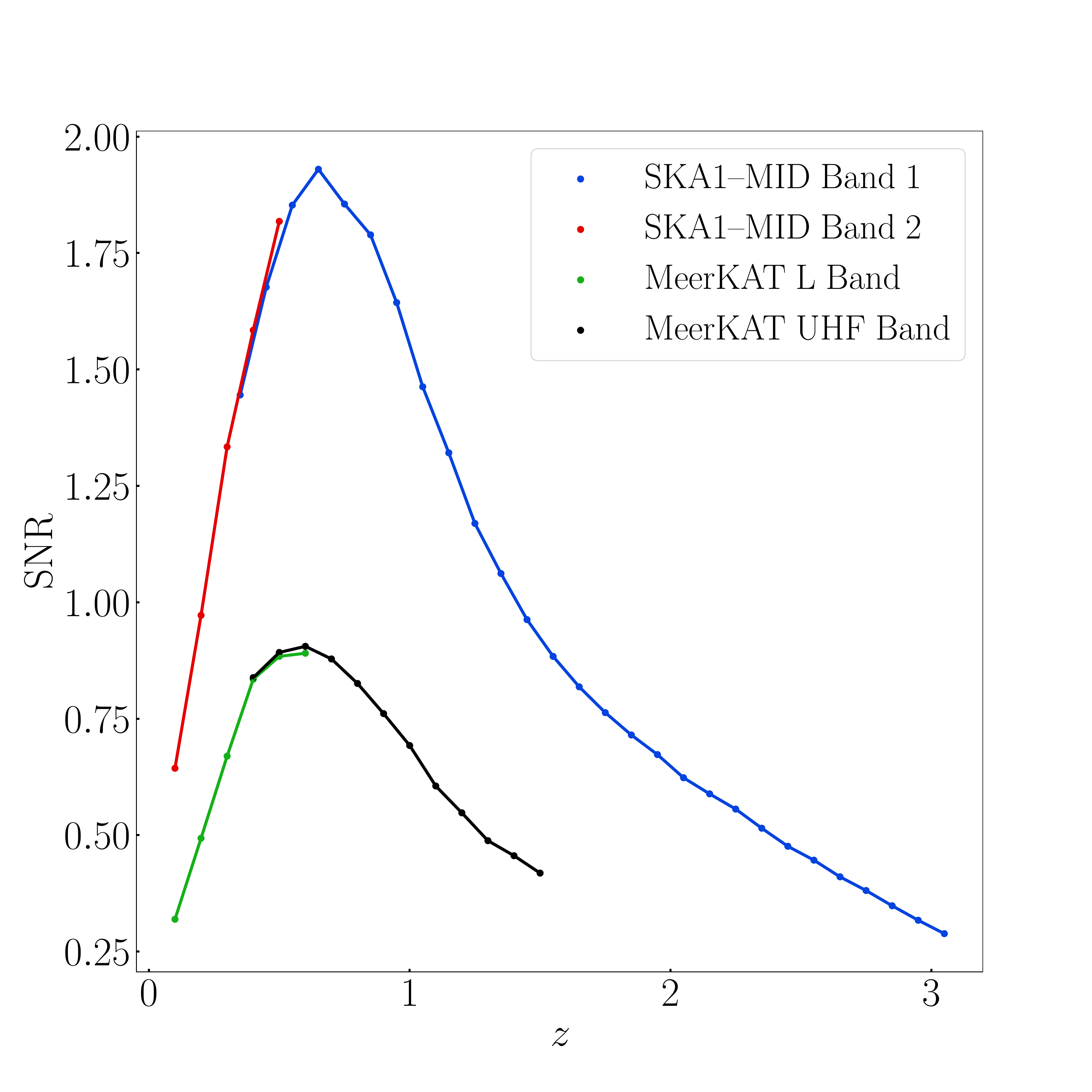}
\includegraphics[width=.49\textwidth]{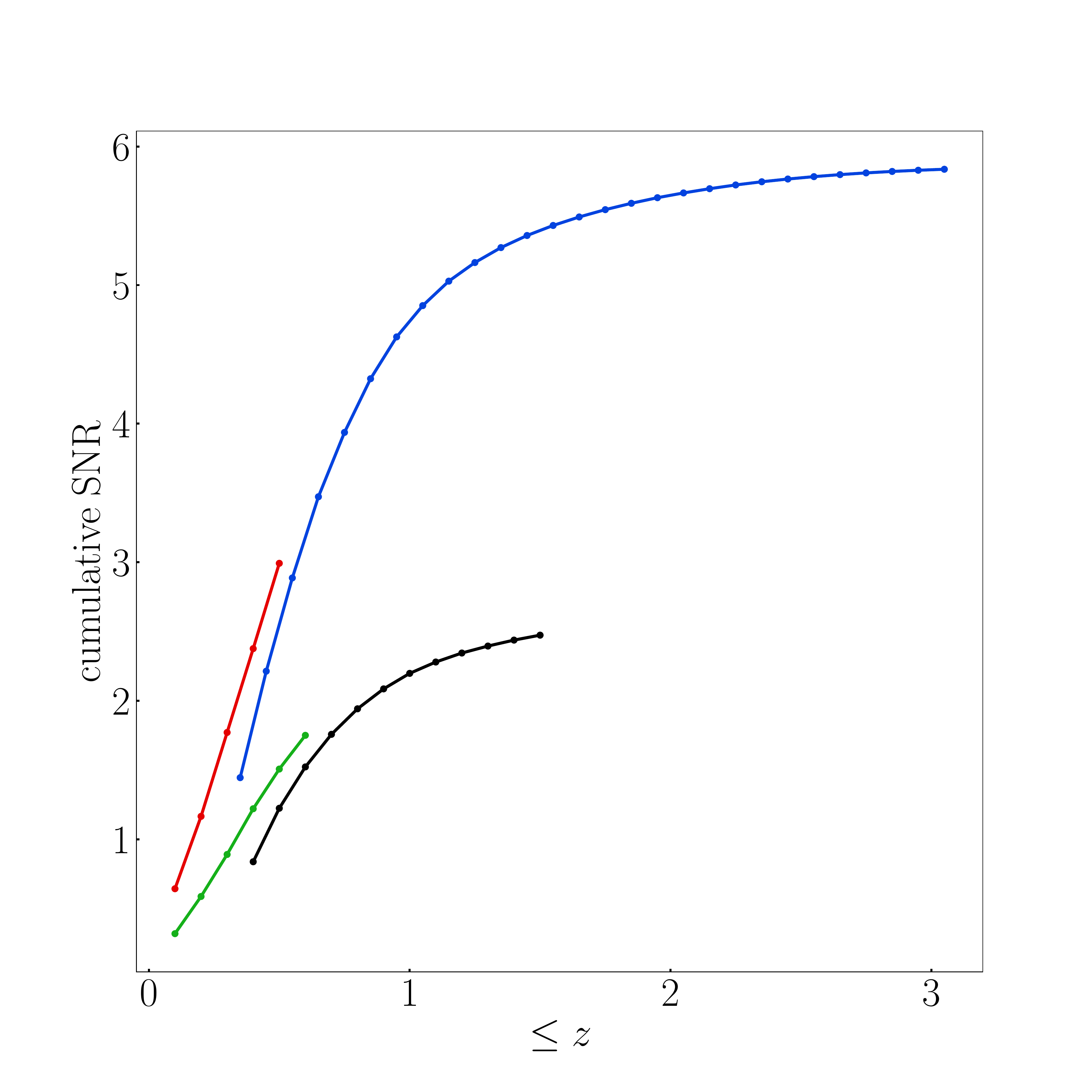} 
\caption{SNR of the relativistic bispectrum per $z$-bin (\emph{left}) and cumulative (\emph{right}) for  SD-mode surveys. }
\label{fig3}
\end{figure}
\begin{table}[h]
\centering
\caption{\label{tab3} Total SNR for single-dish mode surveys.} 
\vspace*{0.2cm}
\begin{tabular}{|lc|} \hline
Survey & Total SNR \\ \hline\hline 
MeerKAT UHF Band & \red{2.5}  \\
MeerKAT L  Band & \red{1.8}  \\
\hline
MeerKAT L+UHF Bands & \red{\bf 3.0}  \\
\hline
SKA1-MID Band 1 & \red{5.8} \\
SKA1-MID Band 2 & \red{3.0} \\
\hline
SKA1-MID Bands 1+2 & \red{\bf 6.6} \\
\hline
\end{tabular}
\end{table} 

The total SNR for the HI IM surveys in SD mode is {$\red{3}\lesssim {\rm SNR} \lesssim \red{7}$, with the high-$z_{\rm max}$ bands giving higher SNR.} 
Table~\ref{tab3} displays the predicted total SNR for the SD-mode HI IM surveys.
We can slightly improve the best total SNR by combining measurements in the two bands, for both MeerKAT and SKA.  In general, since we ignore cross-redshift correlations, we can sum in quadrature the per-bin values, ${\rm SNR}(z_i)$, as if they were a collection of bins from the same set of observations.
However, it must be noted that the two bands overlap in the redshift range  $0.40\leq z\leq 0.58$ for MeerKAT and $0.35\leq z \leq 0.49$ for SKA.
In that range, we  follow a conservative approach and only consider the band yielding the largest value of SNR.
The values obtained are shown in Table~\ref{tab3}.

The best case, SKA1 Bands 1+2, gives a total SNR\,\red{$\sim$6} that is {detectable}, a few times smaller than the SNR predicted for a Stage IV H$\alpha$ (similar to {\em Euclid}) spectroscopic survey \cite{Maartens:2019yhx}. 

\subsection{Interferometer mode}

\begin{figure}[! ht]
\centering
\includegraphics[width=.49\textwidth]{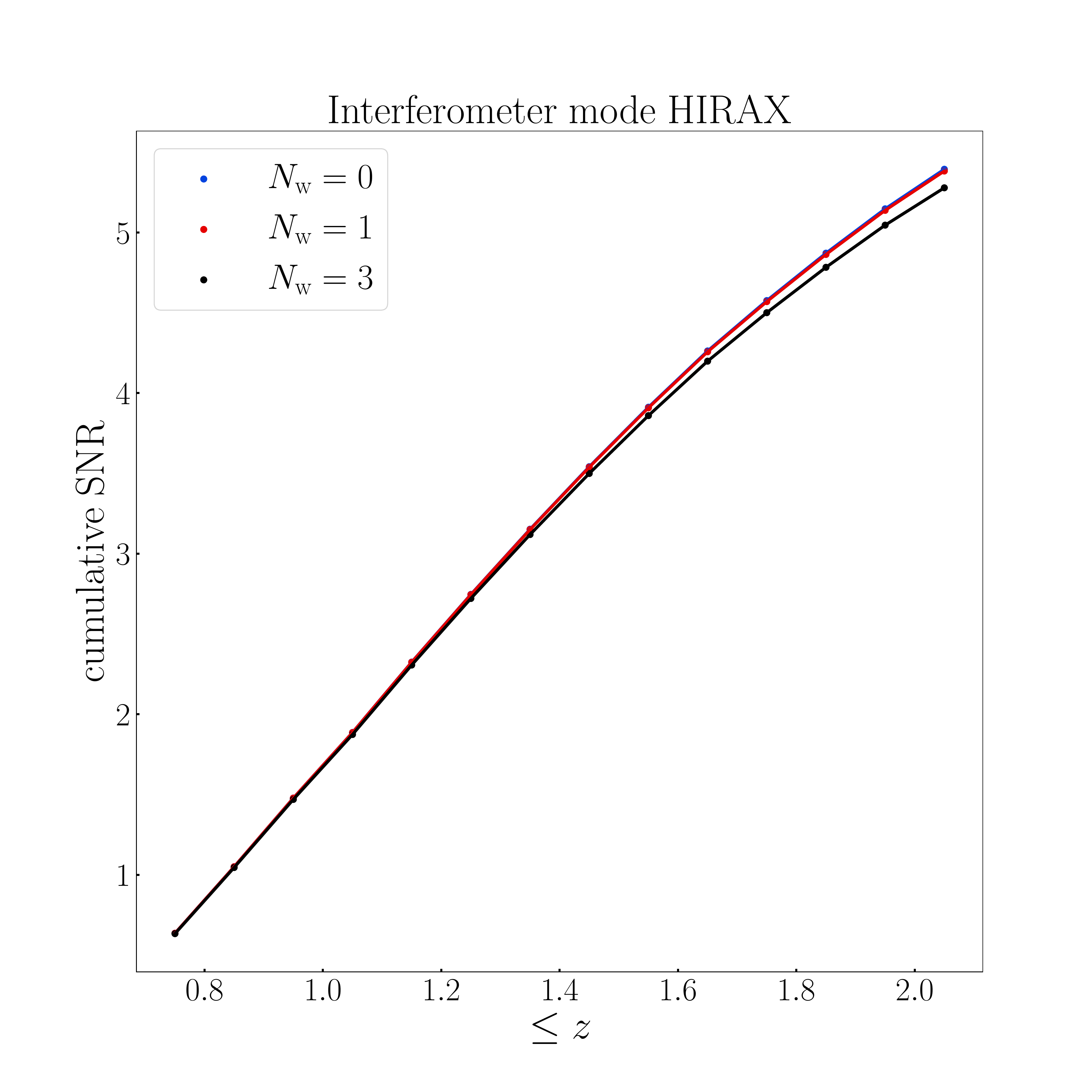} 
\includegraphics[width=.49\textwidth]{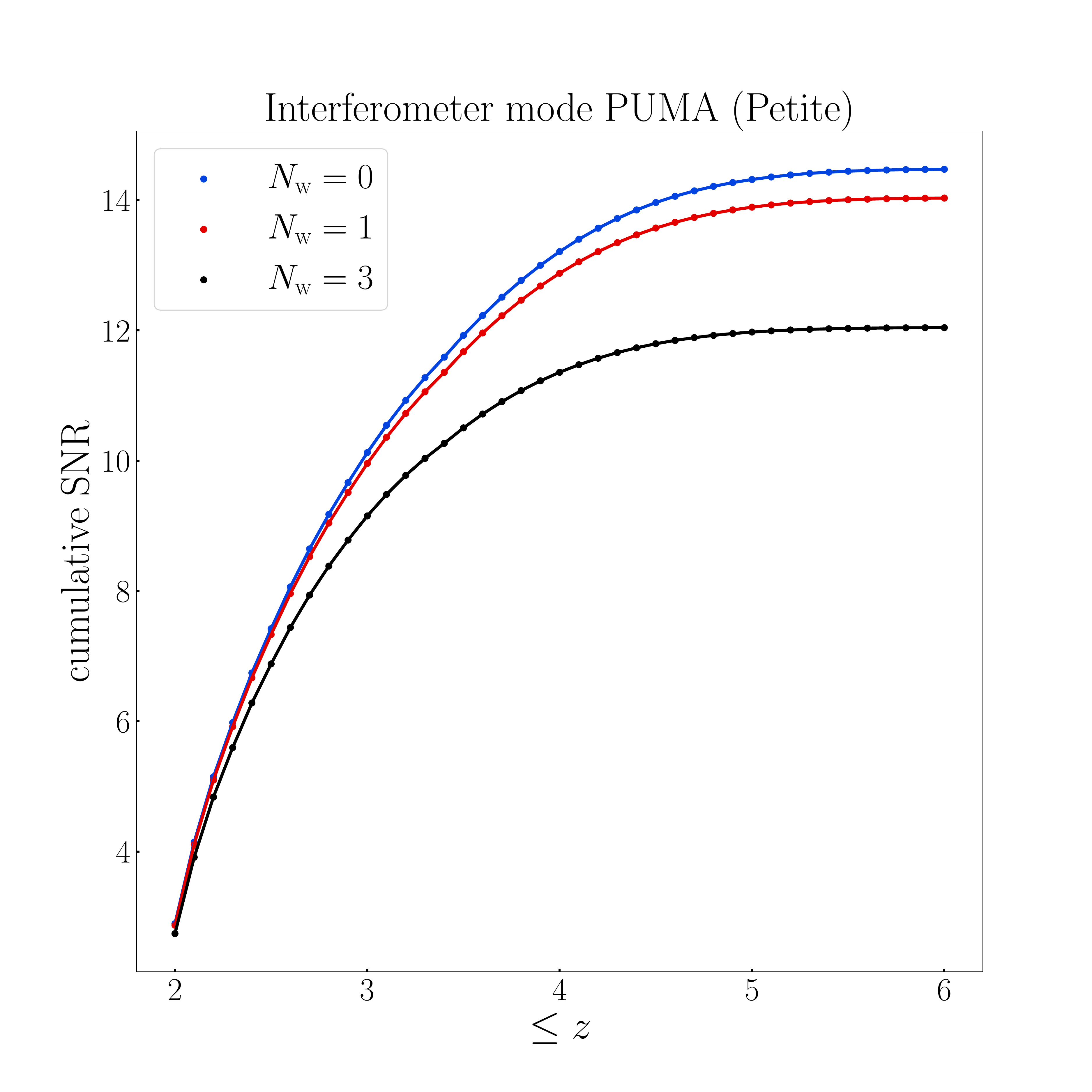}
\caption{Cumulative SNR of the relativistic bispectrum 
for IF-mode surveys.}
\label{fig4}
\end{figure}

\noindent Figure~\ref{fig4} shows the forecasts of the relativistic  bispectrum SNR for  HIRAX and PUMA, using three values of the wedge parameter $N_{\rm w}$. 
HIRAX predicts SNR values {slightly lower than} an SKA SD-mode survey. {The proposed PUMA survey (the `Petite' or PUMA--5K phase of the full or PUMA--32K proposal) gives the highest SNR, at a level similar to a Stage IV H$\alpha$ ({\em Euclid}-like) spectroscopic survey. This SNR would safely detect the relativistic signal.}
Note that PUMA is more sensitive than HIRAX to the $N_{\rm w}$ parameter. 
Table~\ref{tab4} gives the predicted total SNR for the  IF-mode surveys.
\begin{table}[!h]
\centering
\caption{\label{tab4} Total SNR for interferometer-mode surveys.  } 
\vspace*{0.2cm}
\begin{tabular}{|l||*{5}{c|}}\hline
\backslashbox{Survey}{$N_{\mathrm{w}}$}
&\makebox[3em]{0}&\makebox[3em]{1}&\makebox[3em]{3} \\ \hline\hline
HIRAX &\red{5.4} & \red{\bf 5.4} &\red{5.3} \\ \hline
PUMA (Petite) & \red{14.5}& \red{\bf 14.0} & \red{12.0}\\ \hline
\end{tabular} 
\end{table}

\subsection{{Reducing the radial foreground cut}}

{We investigate how much improvement in SNR results if we increase the number of very large scale modes by reducing the radial foreground cut to $k_{\parallel\rm{fg}}=0.005\,h\mathrm{Mpc}^{-1}$. Such a reduction may be achieved by future advances in reconstruction techniques. 

It turns out that the SNR is relatively insensitive to this reduction in $k_{\parallel\rm{fg}}$: very little gain in SNR is achieved, as shown in Tables \ref{tab5a} and \ref{tab6}.}

\begin{table}[!h]
\centering
\caption{\label{tab5a} {Total SNR for single-dish mode surveys with $k_{\parallel\rm{fg}}=0.005\,h\mathrm{Mpc}^{-1}$.} }
\vspace*{0.2cm}
\begin{tabular}{|lc|} \hline
Survey & Total SNR \\ \hline\hline 
MeerKAT UHF Band & \red{2.8}  \\
MeerKAT L Band & \red{2.0}  \\
\hline
MeerKAT L+UHF Bands & {\bf \red{3.4}}  \\
\hline
SKA1-MID Band 1 & \red{6.5} \\
SKA1-MID Band 2 & \red{3.4}\\
\hline
SKA1-MID Bands 1+2 & {\bf \red{7.3}} \\
\hline
\end{tabular}
\end{table} 
\begin{table}[!h]
\centering
\caption{\label{tab6} {Total SNR for interferometer-mode surveys with $k_{\parallel\rm{fg}}=0.005\, h\mathrm{Mpc}^{-1}$.}} 
\vspace*{0.2cm}
\begin{tabular}{|l||*{5}{c|}}\hline
\backslashbox{Survey}{$N_{\mathrm{w}}$}
&\makebox[3em]{0}&\makebox[3em]{1}&\makebox[3em]{3} \\ \hline\hline
HIRAX & \red{5.6} & \red{{\bf 5.6}} & \red{5.5} \\ \hline
PUMA (Petite) & \red{15.2} & \red{{\bf 14.6}} & \red{12.5} \\ \hline
\end{tabular} 
\end{table}


\section{Conclusions} \label{sec5}

{The imaginary part of the galaxy bispectrum at leading order is a unique signal that is purely relativistic \cite{Clarkson:2018dwn} -- and that can be readily extracted from the full bispectrum. In Fourier space, the leading-order local lightcone effects that generate the Doppler-type signal scale as $\mathrm{i}\,\big(\cH/k\big)$ (see \eqref{e1.9} and \eqref{e1.10}). In the galaxy power spectrum for a single tracer, these Doppler-type contributions only occur squared, thus scaling as $\cH^2/k^{2}$. Only a cross-power spectrum analysis of two different tracers will generate an imaginary contribution \cite{McDonald:2009ud}. For single tracers, the relativistic Doppler-type signal in the power spectrum is therefore highly suppressed and not detectable, even for a cosmic-variance limited survey \cite{Alonso:2015uua}.

However, in the bispectrum of a single tracer, the $\mathrm{i}\,\big(\cH/k\big)$ relativistic signal survives because it couples with the Newtonian terms  \cite{Clarkson:2018dwn}. This increases the chance of detectability and, as shown by \cite{Maartens:2019yhx}, the SNR forecast for a Stage IV H$\alpha$ galaxy spectroscopic survey  (similar to {\em Euclid}) is $\sim$17. 

We extended the analysis of  \cite{Maartens:2019yhx} to HI intensity mapping spectroscopic surveys, which requires significant additions in order to deal with foreground contamination and the complexities of instrumental noise (that dominates over shot noise). As expected, loss of signal due to foreground cleaning and telescope beam effects reduces the SNR for the next-generation surveys on MeerKAT, SKA1-MID and HIRAX. 
We forecast a detectable SNR\,\red{$\sim$6} for SKA1 and \red{$\sim$5} for HIRAX.
The proposed PUMA survey in its Petite or 5,000-dish first phase (before the futuristic full survey with 32,000 dishes) delivers a SNR\,\red{$\sim$14}, comparable to that  of a Stage IV H$\alpha$ galaxy survey.

Surveys in single-dish mode (MeerKAT, SKA1) suffer from poor angular resolution so that the Newtonian short-scale transverse modes are suppressed.  On the other hand, the interferometer-mode surveys (HIRAX, PUMA) do not probe ultra-large scales as well as SD-mode surveys because the maximum length scale is determined  by the minimum baseline \eqref{kifmin}, as shown in Figure \ref{pnoise}.}

We investigated the effect of a more optimistic radial wavenumber cut and found that the improvement in SNR is small. {This means that detection of the relativistic signal in the bispectrum does not require scales with  $k\lesssim 0.01h/$Mpc.} 

\red{There are some caveats to our results and pointers for future work: 
\begin{itemize}
\item
In common with nearly all work on the Fourier-space bispectrum that includes RSD, we implicitly make a flat-sky assumption, based on the fixed global direction $\bm n$.   This is an issue when probing ultra-large scales -- which applies not only to our case but also to all work on constraining  primordial non-Gaussianity via the Fourier bispectrum. 
The flat-sky analysis loses accuracy as $\theta$ increases, where $\theta$ is the maximum opening angle to 
the three-point correlations at the given redshift. The corresponding comoving transverse scale is $k_\perp(z) = 2\pi/[r(z) \theta]$. There is a threshold scale  
$\theta_{\rm fs}$, beyond which the  approximation fails -- i.e., for $\theta> \theta_{\rm fs}$, or equivalently, $k_\perp < k_{\perp \rm fs}$, the SNR is not reliable.

For IF-mode  surveys,  the flat-sky assumption is reasonable if  
 $ \theta_{\rm fs}> \theta_{\rm b}$, where $\theta_{\rm b}$ is the beam, given by \eqref{thetab}.  If we estimate that $\theta_{\rm fs} \sim 10^\circ$ (see \cite{Matsubara:1999du}), then  this condition holds for HIRAX, and for PUMA up to redshift 3 (see Figure \ref{figmin}).
Consequently, the flat-sky assumption is reasonable for HIRAX. However for PUMA at $z>3$ and for the SD-mode surveys, there are modes with $k_\perp < k_{\perp \rm fs}$ and the accuracy of the SNR will be affected for these modes.

 Including wide-angle effects   is a key target for future work that constrains relativistic effects or primordial non-Gaussianity via the bispectrum.  In fact, this is also important for standard cosmological constraints.
Corrections to the global flat-sky analysis of  the Fourier bispectrum can be made by using a local flat-sky approximation  \cite{Scoccimarro:2015bla,Sugiyama:2018yzo}.
Such corrections  are typically approximate and do not incorporate the full wide-angle effect. Ultimately, one needs to use the full-sky 3-point correlation function or the full-sky angular bispectrum (see e.g. \cite{Kehagias:2015tda, DiDio:2016gpd, DiDio:2018unb, Durrer:2020orn}) to properly include all wide-angle correlations. These alternatives are computationally much more intensive than the Fourier analysis.

\item 
  Also in common with other work on the Fourier power spectrum and bispectrum, we neglect cross-correlations among redshift bins.  This is justified by the exquisite redshift accuracy of HI intensity mapping, allowing for sharp-edged redshift bins, which in turn implies small or no overlap between them.  Integrated effects will in principle induce correlations along the line-of-sight direction, but this is not going to be relevant in our case since the dominant integrated contribution is weak lensing magnification, which vanishes in HI intensity mapping \cite{Hall:2012wd,Alonso:2015uua,DiDio:2015bua,Jalivand:2018vfz}.    
 Ultimately, only approximate solutions are possible in Fourier space and a complete treatment would require the angular bispectrum or 3-point correlation function.
\item 
In the absence of a simulation-based model for the bispectrum RSD damping parameter $\sigma_B$, we set it equal to the power spectrum damping parameter $\sigma_P$, for which we used a fit from simulations given in \cite{Sarkar:2019nak}. Realistically, we expect that $\sigma_B >\sigma_P$. We tested the impact  on the SNR of increasing $\sigma_B$ to $\sigma_B=1.5\, \sigma_P$ and found that it leads to only a small decrease in SNR. 

\end{itemize}}

\vfill
\noindent {\bf Acknowledgements}\\

\noindent{\small {We thank Emanuele Castorina and Dionysis Karagiannis for helpful comments.}
SJ and RM are supported by  the South African Radio Astronomy Observatory (SARAO) and the National Research Foundation (Grant No. 75415). 
 RM and OU are supported by the UK Science \& Technology Facilities Council (STFC) Consolidated Grant ST/S000550/1. 
CC is supported by STFC Consolidated Grant ST/P000592/1.
SC acknowledges the support from the Ministero degli Affari Esteri della Cooperazione Internazionale - Direzione Generale per la Promozione del Sistema Paese Progetto di Grande Rilevanza ZA18GR02. SC is funded by \textsc{miur} through the Rita Levi Montalcini project `\textsc{prometheus} -- Probing and Relating Observables with Multi-wavelength Experiments to Help Enlightening the Universe's Structure'.} 

\newpage

\appendix

\section{HI bias parameters} \label{app1}

We follow \cite{Umeh:2015gza} to compute $b_{1}$ and $b_{2}$ from a halo-model approach:
\begin{align}
b_{1} &= 1 + \left\langle \frac{2p+\big(q\nu - 1\big)\big[1+(q\nu)^{p}\big]}{\delta_{\rm cr}\big[1+(q\nu)^{p}\big]} \right\rangle _{\!\!\!\rm m}  \!\!, \label{e1.16} \\
b_{2} &= \frac{8}{21}\big(b_{1}-1\big) + \left \langle \frac{2p\big(2p+2\nu q-1\big)+ q\nu\big(q\nu-3\big)\big[1+(q\nu)^{p}\big]}{\delta_{\rm cr}^{2}\big[1+(q\nu)^{p}\big]}  \right \rangle _{\!\!\!\rm m}  \!\!, \label{e1.17}
\end{align}
where the parameters $p$, $q$ and $\nu$ are related to the Sheth-Tormen distribution function (see \cite{Sheth:1999su, Sheth:2001dp, Sheth:1999mn} for more details), and $\delta_{\rm cr}$ is the critical density at which halos collapse spherically \cite{Kitayama:1996ne}, 
\begin{equation}
\delta_{\rm c}(z) = \frac{3(12\pi)^{2/3}}{20}\big[1+0.0123\log{\Omega_{\rm m} (z)}\big] \,. \label{e1.20}
\end{equation}
The mass average is defined by 
\begin{equation}
\big\langle X_{h}(z,\bm{x}) \big\rangle _{\rm m}  = \frac{\int_{M_{-}}^{M_{+}} \ud M\,X_{h}(z,\bm{x},M)\,M_{\mathrm{HI}}(M)\,n_{h}(z,\bm{x},M)}{\int_{M_{-}}^{M_{+}} \ud M\,M_{\mathrm{HI}}(M)\,n_{h}(z,\bm{x},M)} \, , \label{e1.18}
\end{equation}
where $M$ is the mass of {halos} that can host HI gas, and $M_{\pm}$ are the lower and upper mass limits, which are  related to the circular velocities  of the galaxies \citep{Bull:2014rha}. $n_{h}$ is the halo mass function \citep{Sheth:1999su, Sheth:2001dp,Tellarini:2015faa}, and $M_{\rm HI}$ is the HI mass function, which is assumed to follow a power law  \citep{Santos:2015gra},
\begin{equation}
M_{\mathrm{HI}}(M) \propto M^{0.6} \,. \label{e1.19}
\end{equation}
 Figure~\ref{fig1} shows the numerical results from \eqref{e1.16} and \eqref{e1.17}. Fitting formulas for the bias parameters are
\bea
b_{1}(z)& = & 0.754 + 0.0877z + 0.0607z^{2} - 0.00274z^{3}\,, \label{e1.27} \\
b_{2}(z) &= & -0.308 - 0.0724z - 0.0534z^{2} + 0.0247z^{3}\,. \label{e1.28}
\eea
Assuming that  halo formation is a local process in Lagrangian space and that there is no initial tidal bias, the tidal bias is \cite{Tellarini:2015faa}:
\begin{equation}
b_{s^{2}} = \frac{4}{7}\big(1-b_{1}\big)\,.\label{e1.21} 
\end{equation} 

\clearpage

\section{MeerKAT and SKA system temperatures} \label{app2}
\vspace*{-0.5cm}
\begin{table}[! ht]
\centering
\caption{\label{tab5} System temperatures for MeerKAT and SKA1-MID, used in Figure~\ref{fig2} (from \cite{Fonseca:2019qek}).} 
\vspace*{0.2cm}
  \begin{tabular}{|l|l|l|l|l|l|l|l|}
    \hline
      \multicolumn{2}{|c|}{MeerKAT L Band} &
      \multicolumn{2}{c|}{MeerKAT UHF Band} &
      \multicolumn{2}{c|}{SKA1-MID Band 1} & 
      \multicolumn{2}{c|}{SKA1-MID Band 2} \\
      \hline \hline
    $~~~~z~~$ & $~~T_{\rm sys}\;/\;\rm K~~~$ & $~~~z~$ & $~~~~T_{\rm sys}\;/\;\rm K~~$ & $~~~z~$ & $~~T_{\rm sys}\;/\;\rm K~~$ & $~~~~z~~$ & $~~T_{\rm sys}\;/\;\rm K~~$ \\
    \hline
    
  
     0.136 & ~~~~19.2 & 0.420 & ~~~~~~20.3 & 0.403 & ~~~~27.2 & 0.115 & ~~~~16.4 \\
    
     0.183 & ~~~~19.7 & 0.495 & ~~~~~~21.0 & 0.470 & ~~~~26.9 & 0.168 & ~~~~16.6 \\
    
     0.235 & ~~~~20.3 & 0.578 & ~~~~~~21.7 & 0.539 & ~~~~26.8 & 0.223 & ~~~~16.8 \\
    
     0.291 & ~~~~20.9 & 0.671 & ~~~~~~22.5 & 0.612 & ~~~~26.9 & 0.280 & ~~~~17.0 \\
    
     0.352 & ~~~~21.5 & 0.775 & ~~~~~~23.5 & 0.767 & ~~~~27.5 & 0.341 & ~~~~17.2 \\
    
    0.420 & ~~~~22.3 & 0.893 & ~~~~~~24.7  & 0.850 & ~~~~28.1 & 0.403 & ~~~~17.6 \\
    
    0.495 & ~~~~23.1 & 1.03 & ~~~~~~26.1   & 0.938 & ~~~~28.8 & 0.470  & ~~~~18.0 \\
    
    0.578 & ~~~~24.0 & 1.18 & ~~~~~~27.9   & 1.03  & ~~~~29.8 &        &  \\
    
          &          & 1.37 & ~~~~~~30.3   & 1.12  & ~~~~30.8 &        &  \\
     
          &          & 1.45 & ~~~~~~31.5   & 1.22  & ~~~~32.1 &        &  \\
    
          &          &      &              & 1.33  & ~~~~33.5 &        &   \\
    
          &          &      &              & 1.44  & ~~~~35.2 &        &    \\
    
          &          &      &              & 1.55  & ~~~~37.1 &        &   \\
    
          &          &      &              & 1.67  & ~~~~39.2 &        &   \\
    
          &          &      &              & 1.80  & ~~~~41.6 &        & \\
    
          &          &      &              & 1.93  & ~~~~44.2 &        &  \\
    
          &          &      &              & 2.07  & ~~~~47.2 &        &  \\
    
          &          &      &              & 2.22  & ~~~~50.6 &        &  \\
    
          &          &      &              & 2.37  & ~~~~54.4 &        &  \\
    
          &          &      &              & 2.54  & ~~~~58.6 &        &  \\
    
          &          &      &              & 2.69  & ~~~~63.4 &        &  \\
    
          &          &      &              & 2.87  & ~~~~68.8 &        &  \\
    
          &          &      &              & 3.05  & ~~~~74.8 &        &  \\
    
    \hline
  \end{tabular}
\end{table}
\vspace*{-0.5cm}
 \begin{figure}[! h]
\centering
\includegraphics[width=.49\textwidth]{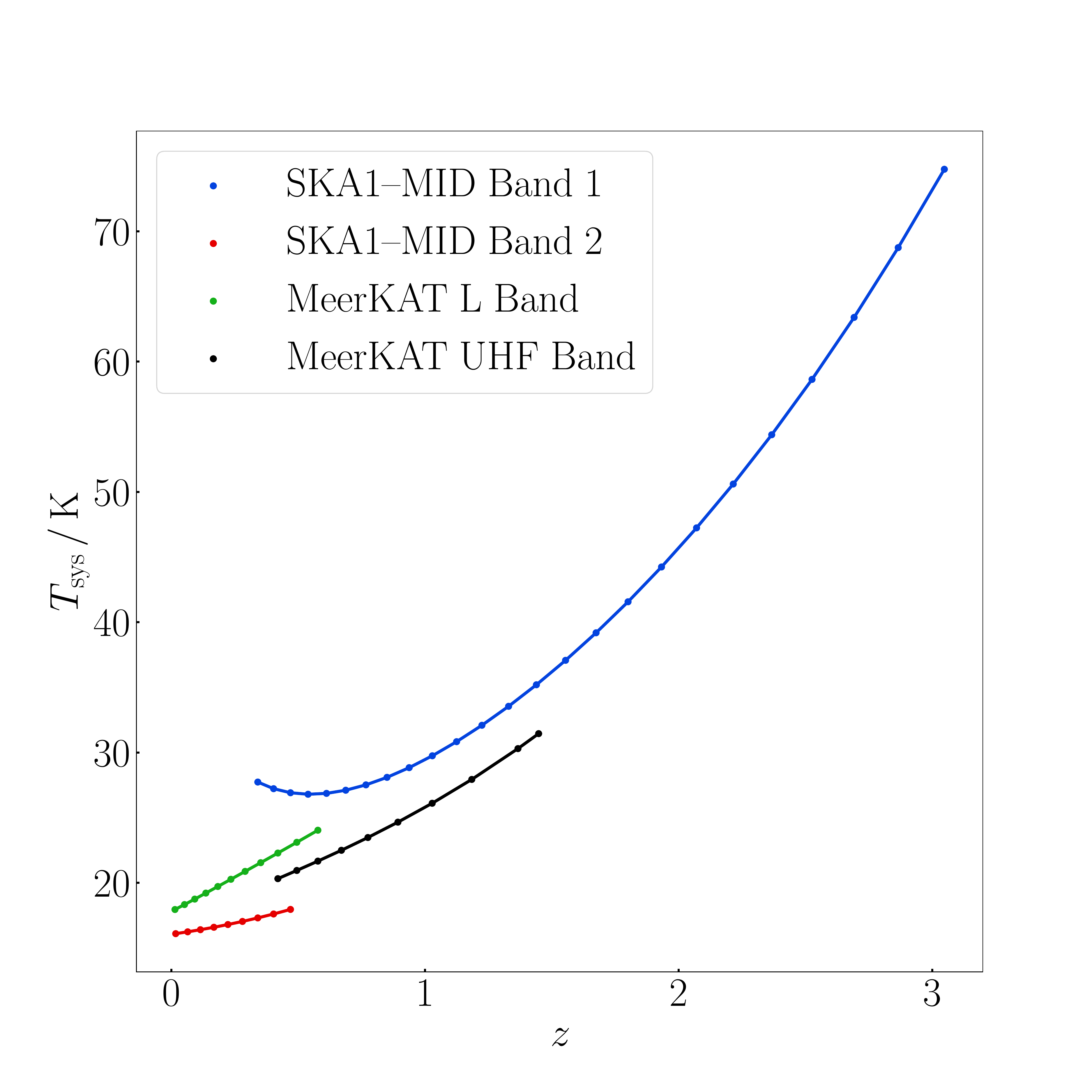}
\vspace*{-0.5cm}
\caption{$T_{\rm sys}$ for the different frequency bands of MeerKAT and SKA1 (from Table \ref{tab5}).} \label{fig2}
\end{figure}

\clearpage
\bibliographystyle{JHEP}
\bibliography{reference_library}

\end{document}